\documentclass[a4paper,11pt]{article}
\pdfoutput=1 
\usepackage{jheppub} 
\usepackage[T1]{fontenc} 
\def\1{\mathbf{1}}
\def\00{{\mathbf 0}}
\newcommand{\Aa}{\mathcal{A}}
\def\CC{{\mathbb C}}
\def\HH{{\mathbb H}}
\def\ZZ{{\mathbb Z}}

\title{\boldmath Spectral action and the electroweak $\theta$-terms for the Standard Model without fermion doubling}

\author{A.~Bochniak,}
\author{A.~Sitarz}
\author{and P.~Zalecki}
\affiliation{Institute of Theoretical Physics, Jagiellonian University,	\\
	prof.\ Stanis\l awa \L ojasiewicza 11, 30-348 Krak\'ow, Poland.}

\emailAdd{arkadiusz.bochniak@doctoral.uj.edu.pl}
\emailAdd{andrzej.sitarz@uj.edu.pl}
\emailAdd{pawel.zalecki@doctoral.uj.edu.pl}

\abstract{We compute the leading terms of the spectral action for a noncommutative geometry model 
that has no fermion doubling. The spectral triple describing it, which is chiral and allows for CP-symmetry breaking, 
has the Dirac operator that is not of the product type. Using Wick rotation we  derive explicitly the Lagrangian 
of the model from the spectral action for a flat metric, demonstrating the appearance of the topological
$\theta$-terms for the electroweak gauge fields.}
\begin{document} 
\maketitle
\flushbottom

\section{Introduction}

The Standard Model of Particle Physics is a powerful theory that gives a precise and effective description
of all fundamental forces apart from gravity. Its predictive power and agreement with experimental results 
guarantee that it needs to remain the backbone of any fundamental theory of particle interactions. Yet, 
in contrast to General Relativity, which is deeply rooted in geometry of the space-time, the Standard Model only
partially can be explained in a similar manner. The structure of gauge theory and the Yang-Mills action 
signifies that indeed the geometry plays there a significant role. However, the appearance of a Higgs field and
the symmetry-breaking quartic potential are not directly implied by the classical geometry.

The hint that the Standard Model has more subtle structure came from noncommutative geometry and the 
theory of spectral triples. Founded by Alain Connes to solve significant mathematical problems related to the 
index, the theory is a well-structured non-trivial generalization of classical differential geometry that allows to 
study not only differentiable manifolds but also discrete spaces, fractals and quantum deformations of 
spaces from a novel point of view. Interestingly, the tools of noncommutative geometry allowed to construct
models that explain the geometry of the Standard Model \cite{Co95,CoLo91, Co96} (see also \cite{CoBook} 
and \cite{walter} for detailed discussion) and its extensions \cite{stephan1,stephan2,stephan3,lizziGRAND,lizziGRAND1}. 
Their structure is similar to Kaluza-Klein models yet with a finite noncommutative algebra instead of the additional 
dimension of space-time. The geometry of the entire enhanced space-time is determined by a Dirac operator that depends 
on the metric and the gauge connections, and also includes the Higgs field, which plays a role of a connection 
over the finite noncommutative component. The spectral action then gives the full gravity and Yang-Mills action 
with the quartic Higgs potential and minimal couplings between the Higgs and the gauge fields  \cite{CCM07}. 

The story of the noncommutative model-building is, however, not yet complete as the mostly accepted model is
in the Euclidean signature and requires  additional assumptions to remove the possibility of the $SU(3)$ 
symmetry breaking \cite{PSS99, DDS18} as well as additional projection onto the physical space of fermions (due to the fermion 
quadrupling in the model) \cite{LMMG97, GBIS98, DKL16}. In the analysis of the Lorentzian case with a slight modifications of the 
spectral triple rules we proved that there exists a model without the fermion doubling and with exact colour $SU(3)$ 
symmetry \cite{BS20}.  Moreover, the non-product Dirac operator satisfied a slightly modified 
first-order condition which is tantamount to spin${}_c$ condition under certain conditions on mass spectra of fermions. 
The CP-symmetry breaking in the Standard Model was then geometrically explained as the lack of reality symmetry of 
the finite component of the Dirac operator as witnessed by nonvanishing of the Wolfenstein parameter and the CP-phase 
in the neutrino sector.

In the paper we compute the spectral action for the model we presented in \cite{BS20}. It needs to be stressed that 
this model is not of the product-type geometry and therefore the computations and results are not automatically 
identical  to these performed in the series of papers computing the spectral action \cite{CCM07, connes_new}. In addition, 
as we start with the Lorentzian model we need to perform a Wick rotation to be able to use heat trace techniques 
\cite{V_manual} or restrict the model to spatial and time-independent (static) components of the fields. To check 
the consistency of the computations we perform both operations. The new element of the spectral action, apart 
from slight differences in the relative coefficients, is the appearance of topological theta terms for the gauge fields 
in the electroweak sector. This is a characteristic new feature of this model, which is inherently chiral, especially
that such terms cannot appear in the spectral action of the product geometries. 

\section{The starting point: fermions and the algebra of the Standard Model}

We begin by briefly reviewing the model as described in details in \cite{BS18, BS20}.  The particle content 
in the one-generation Standard Model can be conveniently parametrized in the following form:
\begin{equation}
\Psi=\begin{pmatrix}
\nu_R & u_R^1 & u_R^2 & u_R^3 \\
 e_R & d_R^1 & d_R^2 & d_R^3 \\
\nu_L & u_L^1 & u_L^2 & u_L^3 \\
 e_L & d_L^1 & d_L^2 & d_L^3
\end{pmatrix} \in M_4(H_W).
\end{equation}
Every entry of the above matrix is a Weyl spinor (from $H_W$) over the Minkowski space $\mathcal{M}^{1,3}$. The algebra 
$\Aa$ is taken to consist of (smooth)  $ \CC \oplus \HH \oplus M_3(\CC)$-valued functions over $\mathcal{M}^{1,3}$. 
We choose its left and right real representations:
\begin{equation*}
\pi_L(\lambda, q ,m ) \Psi = 
\left( \begin{array}{ccc} \lambda & & \\ & \bar{\lambda} & \\ & & q \end{array}\right) \Psi ,
\qquad
\pi_R(\lambda, q ,m ) \Psi = 
\Psi \begin{pmatrix}
\bar{\lambda} &  \\
& m^\dagger
\end{pmatrix},   
\end{equation*}
where $\lambda, q$ and $m$ are complex, quaternion and $M_3(\CC)$-valued functions, respectively. Since left and right 
multiplications commute, the zeroth-order condition is satisfied, i.e. 
\begin{equation*}
[\pi_L(a),\pi_R(b)]=0
\end{equation*} for all $a,b\in \Aa$. It is convenient to encode local linear operator acting on the particle content of the model, 
at every point of $\mathcal{M}^{1,3}$, as an element of $M_4(\CC) \otimes M_2(\CC)\otimes M_4(\CC)$, where the first and 
the last matrices act by multiplication from the left and from the right, respectively, while the middle $M_2(\CC)$ matrix acts 
on the components of the Weyl spinor.

Using this notation, the full Lorentzian Dirac operator of the Standard Model can be written of the form,
\begin{equation}
D_{SM} \Psi = \underbrace{
\begin{pmatrix}
&& i\widetilde{\sigma}^\mu\partial_\mu &\\
&&& i\widetilde{\sigma}^\mu \partial_\mu \\
i\sigma^\mu \partial_\mu &&&\\
& i \sigma^\mu \partial_\mu &&
\end{pmatrix}}_{D}\Psi + D_F\Psi,
\end{equation}
where $\sigma^0 = \1_2 = \widetilde{\sigma}^0$ and $\widetilde{\sigma}^i = - \sigma^i$, the latter being standard Pauli matrices.
$D_F$ is a finite endomorphism of the Hilbert space $M_4(H_W)$.

In \cite{BS20} the Krein-shifted full Dirac operator of the Standard Model,  $\widetilde{D_{SM}} = \beta D_{SM}$, where 
\begin{equation}
\beta =\begin{pmatrix}	& \1_2   \\ \1_2 & \end{pmatrix}\otimes \1_2\otimes \1_4,
\end{equation}
was discussed in details. The Krein-shifted manifold component of the Lorentzian Dirac operator $\widetilde{D}$ in the local 
Cartesian coordinates over $\mathbb{R}^4$,  with a flat metric, is
\begin{equation}
\begin{split}
\widetilde{D} =
\begin{pmatrix}
	\1_2 &  \\ & \00_2
\end{pmatrix}   \otimes i\sigma^\mu \partial_\mu\otimes \1_4 
+ 
\begin{pmatrix}
	\00_2 &  \\ & \1_2
\end{pmatrix}   \otimes i \widetilde{\sigma}^\mu \partial_\mu\otimes \1_4,
\end{split}
\end{equation}
whereas the Krein-shifted discrete part of the Dirac operator is,
\begin{equation}
\label{DFfull}
\widetilde{D_F} = 
\underbrace{\begin{pmatrix}
& M_l \\ M_l^\dagger &
\end{pmatrix}}_{D_l}\otimes \1_2 \otimes e_{11} +
\underbrace{\begin{pmatrix}
	& M_q \\ M_q^\dagger &
\end{pmatrix}}_{D_q} \otimes \1_2 \otimes (\1_4 - e_{11}),
\end{equation}
where $M_l,M_q\in M_2(\mathbb{C})$.

The Krein-shifted Dirac operator and the algebra were proven to satisfy the modified order-one condition: for all $a,b \in A$,
\begin{equation}
	\left[ \pi_R(a), [\widetilde{D_{SM}}, \pi_L(b)] \right] =0, \qquad 	\left[ \pi_L(a), [\widetilde{D_{SM}}, \pi_R(b)] \right] =0.
\end{equation} 

The Lorentzian spectral triple for the signature $(1,3)$ has the standard chirality $\ZZ_2$-grading $\gamma$ and the 
charge conjugation operator,
$\mathcal{J}$, 
\begin{equation}
	\gamma = 
	\begin{pmatrix}
		1_2 & 0 \\
		0 & -1_2
	\end{pmatrix}, 
	\qquad
	\mathcal{J} =i\gamma^2\circ cc
	= i \begin{pmatrix}
		0 & \sigma^2\\
		-\sigma^2 & 0
	\end{pmatrix} \circ cc,
\end{equation}
where $cc$ denotes the usual complex conjugation of spinors. The construction can be easily generalized for the three families 
of leptons and quarks by enlarging  the Hilbert space diagonally, i.e. by taking $M_4(H_W) \otimes \CC^3$ with the diagonal 
representation of the algebra. The matrices $M_l$ and $M_q$ in \eqref{DFfull} 
are no longer in $M_2(\CC)$ but in $M_2(\CC) \otimes M_3(\CC)$.  Its standard presentation for the physical Standard Model is 
\begin{equation*}
M_l = \begin{pmatrix} \Upsilon_\nu & 0 \\ 0 & \Upsilon_e \end{pmatrix},
\qquad
M_q= \begin{pmatrix} \Upsilon_u & 0 \\ 0 & \Upsilon_d \end{pmatrix},    
\end{equation*}
where $\Upsilon_e$ and $\Upsilon_u$ are chosen diagonal with the masses 
of electron, muon, and tau and the up, charm, and top quarks, respectively, and $\Upsilon_\nu$ and $\Upsilon_d$ can be diagonalised by the Pontecorvo-Maki-Nakagawa-Sakata mixing matrix (PMNS matrix) $U$ and the Cabibbo-Kobayashi-Maskawa mixing matrix (CKM matrix) $V$, respectively:
\begin{equation*}
\Upsilon_\nu = U \widetilde{ \Upsilon_\nu} U^\dagger,  \qquad
\Upsilon_d = V \widetilde{ \Upsilon_d} V^\dagger.    
\end{equation*}
The matrices $\widetilde{ \Upsilon_\nu}, \widetilde{ \Upsilon_d}$ provide
(Dirac) masses of all neutrinos and down, strange, and bottom quarks.

As it was demonstrated in \cite{BS20} the model has interesting algebraic properties, like the Morita duality (which means 
that the generalized Clifford algebra is a full commutant of the algebra $\mathcal{A}$) provided that both pairs of 
matrices $(\Upsilon_\nu,\Upsilon_e)$ and $(\Upsilon_u,\Upsilon_d)$ have pairwise different  eigenvalues. Furthermore, 
without referring to additional symmetries or assumptions the model preserves the $SU(3)$ strong symmetry 
and allows for the natural breaking of the CP-symmetry, which is linked to the non-reality of the mixing matrices. 
This is, on the level of algebra of the model, equivalent to the failure of the finite part of the Krein-shifted Dirac operator 
to be ${\mathcal{J}}$-real (see \cite{BS20} for details).

\subsection{The gauge transformations and the Higgs}

In this section we extend the model by intruding the fluctuations of the Dirac operator, that is a family of operators obtained 
from $\widetilde{D_{SM}}$ by perturbing them with one forms, that originate from commutators with the Dirac operator.  
Here, both left and right representations will contribute to the gauge transformations and to the fluctuations of the Dirac 
operator. 

A gauge transformation of physical fields is a unitary transformation of the form,
\begin{equation}
U_{LR}=\pi_L(U) \pi_R(U), 
\end{equation} 
for a unitary element $U$ of the algebra $\mathcal{A}$, so that the gauge-transformed Dirac operator becomes:
\begin{equation}
	\widetilde{D_{SM}}^U = \pi_L(U) \pi_R(U) \widetilde{D_{SM}}  \pi_R(U^\dagger) \pi_L(U^\dagger),
\end{equation}
which, after using the order-zero and order-one conditions, can be rewritten as
\begin{equation}
	\widetilde{D_{SM}}^U = \widetilde{D_{SM}}  + 
	 \pi_L(U) \bigl[ \widetilde{D_{SM}},   \pi_L(U^\dagger) \bigr] + 
	 \pi_R(U) \bigl[ \widetilde{D_{SM}} , \pi_R(U^\dagger) \bigr].
\end{equation}
The unitary group of the finite algebra is $U(1)\times SU(2)\times U(3)$, while the unitaries of the form $U_{LR}$ are elements of the group being a 
quotient of this group by its diagonal normal subgroup $\mathbb{Z}_2=\{\pm(1, \1_2,\1_3)\}$. 

To parametrize the fluctuations we first start with computing left and right ones separately:
\begin{equation}
\sum\limits_{j}\pi_L(a_j)[\widetilde{D_{SM}},\pi_L(b_j)], \qquad \sum\limits_{j}\pi_R(a_j)[\widetilde{D_{SM}},\pi_R(b_j)],
\end{equation}
where $a_j,b_j\in \mathcal{A}=C^\infty (\mathbb{R}^4,\CC \oplus \HH \oplus M_3(\CC))$, and the representations 
$\pi_L$ and $\pi_R$ are of the form:
\begin{equation}
\pi_L(a)=\underbrace{\begin{pmatrix}
\lambda_a &&\\
& \overline{\lambda_a}\\
&& q_a\end{pmatrix}}_{a^L}\otimes \,\mathbf{1}_2\otimes \mathbf{1}_4, \qquad \pi_R(a)=\mathbf{1}_4\otimes \1_2\otimes \underbrace{\begin{pmatrix}
\overline{\lambda_a} & \\
& m^\dagger_a
\end{pmatrix}}_{a^R}
\end{equation}
where $\lambda_a\in C^\infty(\mathbb{R}^4),\ q_a\in C^\infty(\mathbb{R}^4,\mathbb{H})$ and $m_a\in C^\infty(\mathbb{R}^4, M_3(\mathbb{C}))$. 

We first notice that $[\widetilde{D_F},\pi_R(b)]=0$ from the very definition of the representation and the structure of this Dirac operator. Therefore, the only contribution from the right fluctuations can be parametrized as
\begin{equation}
\begin{pmatrix}	\1_2 &  \\ & \00_2 \end{pmatrix}  
              \otimes \sigma^\mu \otimes    
             \begin{pmatrix}   A'_\mu &  \\ & G_\mu  \end{pmatrix} 
           +     \begin{pmatrix}	\00_2 &  \\ & \1_2   \end{pmatrix}   
            \otimes \widetilde{\sigma}^\mu\otimes
       \begin{pmatrix}   A'_\mu &  \\ & G_\mu  \end{pmatrix}, 
\end{equation}
where $A_\mu'=i\sum\limits_{j}\overline{\lambda_{a_j}}(\partial_\mu\overline{\lambda_{b_j}})$ and $G_\mu=i\sum\limits_{j}m_{a_j}^\dagger\left( \partial_\mu m_{b_{j}}^\dagger\right)$.

Now, we compute the left fluctuations. Starting with the ones following from the $\widetilde{D}$ part of the Dirac operator we immediately get
\begin{equation}
\sum\limits_{j}\pi_L(a_j)[\widetilde{D},\pi_L(b_j)]=A_\mu^R\otimes \sigma^\mu \otimes \1_4 + A_\mu^L \otimes\widetilde{\sigma}^\mu\otimes \1_4,
\end{equation}
with
\begin{equation}
A_\mu^R=\begin{pmatrix}
A_\mu&&\\ & A_\mu' & \\ && 0_2
\end{pmatrix}, \quad
A_\mu^L=\begin{pmatrix}
0_2&\\ & W_\mu
\end{pmatrix},
\end{equation}
where $A_\mu=i\sum\limits_{j}\lambda_{a_j}(\partial_\mu\lambda_{b_j})$, $A_\mu'$ is as previously, and $W_\mu=i\sum\limits_{j} q_{a_j}(\partial_\mu q_{b_j})$.

Imposing the selfadjointness condition we immediately get $A_\mu'=-A_\mu$ and infer that $W_\mu$ is indeed an element of $i\mathfrak{su}(2)$ (as it is enforced to be a real linear combination of Pauli matrices). Similarly, we deduce that $G_\mu$ is a $U(3)$ gauge field.

It remains to take into account the contribution from $\widetilde{D_F}$. By a straightforward computation we get
\begin{equation}
\sum\limits_{j}\pi_L(a_j)[\widetilde{D_F},\pi_L(b_j)]=\phi^{l}\otimes 1_2\otimes e_{11} + \phi^{q}\otimes 1_2\otimes (1_4-e_{11}),
\end{equation}
where 
\begin{equation}
\phi^{r}=\sum\limits_j a_j^L[D_r,b_j^L],\qquad r=l,q.
\end{equation}

Since both matrices $M_l$ and $M_q$ are diagonal, they commute with $\mathrm{diag}(\lambda,\overline{\lambda})$. Denoting by 
\begin{equation*}
    \mathbf{\Phi}=\sum\limits_{j}\begin{pmatrix}
\lambda_{a_j}&\\ &\overline{\lambda_{a_j}}
\end{pmatrix}\left[ q_{b_j} - \begin{pmatrix}
\lambda_{b_j}& \\
& \overline{\lambda_{b_j}}
\end{pmatrix} \right],
\end{equation*}
we can parametrize those fluctuations, under the assumption of selfadjointness, as:
\begin{equation}
 \begin{pmatrix}
        	& M_l {\mathbf \Phi}  \\ {\mathbf \Phi}^\dagger  M_l^\dagger &
               \end{pmatrix} \otimes \1_2 \otimes e_{11} +
              \begin{pmatrix}
	        & M_q {\mathbf \Phi}  \\ {\mathbf \Phi}^\dagger  M_q^\dagger &
               \end{pmatrix} \otimes \1_2 \otimes (\1_4 - e_{11}).
\end{equation}

As a result, the fluctuations coming from all the terms can be parametrize as
\begin{equation}
	\begin{split}
         \omega&=A_\mu e_{11} \otimes \sigma^\mu  \otimes (\1_4 - e_{11})
            - 2 A_\mu e_{22} \otimes \sigma^\mu  \otimes e_{11}  \\
        & - A_\mu e_{22} \otimes \sigma^\mu  \otimes (\1_4 - e_{11}) 
            - A_\mu (e_{33}+e_{44}) \otimes \tilde{\sigma}^\mu  \otimes e_{11} \\
        & + \begin{pmatrix}
        	\00_2 &  \\ & W_\mu \end{pmatrix}
           \otimes \widetilde{\sigma}^\mu  \otimes \1_4  \\
        & + \begin{pmatrix}	\1_2 &  \\ & \00_2 \end{pmatrix}  
              \otimes \sigma^\mu \otimes    
             \begin{pmatrix}   \00_1 &  \\ & G_\mu  \end{pmatrix} 
           +     \begin{pmatrix}	\00_2 &  \\ & \1_2   \end{pmatrix}   
            \otimes \widetilde{\sigma}^\mu\otimes
       \begin{pmatrix}   \00_1 &  \\ & G_\mu  \end{pmatrix} \\
        & +  \begin{pmatrix}
        	& M_l {\mathbf \Phi}  \\ {\mathbf \Phi}^\dagger  M_l^\dagger &
               \end{pmatrix} \otimes \1_2 \otimes e_{11} +
              \begin{pmatrix}
	        & M_q {\mathbf \Phi}  \\ {\mathbf \Phi}^\dagger  M_q^\dagger &
               \end{pmatrix} \otimes \1_2 \otimes (\1_4 - e_{11}).
\end{split}
\end{equation}
We denote the fluctuated Dirac operator by $\widetilde{D_{SM}}^\omega=\widetilde{D_{SM}}+\omega$.

For a unitary element $U\equiv (u_1, u_2, u_3) \in U(1) \times SU(2) \times U(3)$
the gauge-transformed fluctuated Dirac operator is of the form
\begin{equation}
\widetilde{D_{SM}}^{\omega^U}=\pi_L(U) \pi_R(U) \widetilde{D_{SM}}^\omega  \pi_R(U^\dagger) \pi_L(U^\dagger).
\end{equation}
The gauge transformation can be therefore implemented by
\begin{equation*}
    \begin{split}
        \omega\longmapsto \omega^U&=\pi_L(U) \pi_R(U) \omega  \pi_R(U^\dagger) \pi_L(U^\dagger)\\ &+ \pi_L(U) \left[ \widetilde{D_{SM}},   \pi_L(U^\dagger) \right] + 
	 \pi_R(U) \left[ \widetilde{D_{SM}} , \pi_R(U^\dagger) \right].
    \end{split}
\end{equation*}
As a result, the fields $A_\mu,W_\mu,G_\mu, {\mathbf \Phi}$ transform accordingly:
\begin{equation}
\begin{split}
& A_\mu \longmapsto A_\mu + u_1 (\partial_\mu \overline{u_1}), \\
&  W_\mu \longmapsto u_2W_\mu u_2^\dagger + u_2 (\partial_\mu u_2^\dagger), \\
& G_\mu \longmapsto u_3G_\mu u_3^\dagger+u_3 (\partial_\mu u_3^\dagger),\\
& \1_2+\mathbf{\Phi} \longmapsto \begin{pmatrix}  u_1 &  \\ &  \overline{u_1} \end{pmatrix}  (\1_2+{\mathbf \Phi})  u_2^\dagger.
\end{split}
\end{equation}

We remark that in the above derivation the crucial role was played by the fact that the representation of $U(1)$ part of the 
gauge group commutes with the mass and mixing matrices.

It is known that the gauge group of the Standard Model should contain $SU(3)$ rather than $U(3)$. This can be achieved with a further condition,
the unimodularity of the representation, which, however, can be imposed in different ways. In particular, let us observe that the left action of the
group is unimodular from the beginning. For the right representation one could require either the condition that it is unimodular on each fundamental
component (chiral lepton and quark) or in the full representation, including all chiral fermions and families. In the first case, the unimodularity 
condition is equivalent to $u_1\det u_3=1$, whereas in the second case it becomes $(u_1\det u_3)^{12}=1$.  In the first case, the resulting group 
is exactly the group of the Standard Model,
\begin{equation*}
    \left(U(1)\times SU(2)\times SU(3)\right)/\mathbb{Z}_6,
\end{equation*}
whereas in the latter case it is the one described in \cite[Prop.~11.4]{walter}, which differs from the gauge group of the Standard Model 
by a finite factor. Independently, their Lie algebras agree and the finite difference does not affect the structure of the gauge fields. 
The unimodularity condition on the algebraic level of perturbation means that the trace of a perturbation has to vanish. 
This condition results in $\mathrm{Tr}(G_\mu)=A_\mu$. We therefore 
introduce the traceless gauge field $G_\mu'=G_\mu-\frac{1}{3}A_\mu \1_3$ and then in the perturbations we can replace 
$G_\mu$ by $G_\mu+\frac{1}{3}A_\mu \1_3$, where now $G_\mu$ is assumed to be a $SU(3)$ gauge field. 

\section{The spectral action}

In the considerations so far (see e.g. \cite{walter} and references therein), the spectral action for the Standard Model was computed 
for the Euclidean model with fermion doubling. Moreover, the assumed Dirac operator was of the product type and therefore its square 
was simply the sum of the squares of the Dirac operators on manifold component and on the discrete component.  While this strategy 
can be justified by the arguments of covariance and geometric character of the action, the relative coefficients and the proportions 
between them may in general differ, when one considers the Lorentzian and explicitly chiral Dirac operator. 

Of course, the best strategy would be to apply genuine Lorentzian approach (see \cite{Wr21}), however, this appear to be at 
the moment restricted only to scalar operators and not Dirac-type operators. Therefore we propose two simple, computable 
methods to obtain an insight into the action of the model, which is motivated by spectral methods.

The first one assumes that we restrict ourselves to the {\bf static and spatial} case, computing the terms of the spectral 
action for the Krein-shifted Dirac operator that is restricted to the spatial part and with the gauge fields that arise exclusively 
through static (time independent) gauge transformations. Such restricted Dirac operator is already a hermitian elliptic operator 
and one can easily compute the heat trace coefficients of its square. This shall recover the action of the model 
for the time-independent fields, which is invariant under static gauge transformations. However, one cannot expect that all 
terms of the action will be present, and their coefficients correct. 

The second method takes as the input the {\bf Wick-rotated Lorentzian Dirac} operator $D_w$.  Such operator is elliptic, as its 
continuous part is just the usual Wick-rotated Dirac  operator  (with gauge fields) over the flat space-time. However, the 
discrete part of the operator (which is not Krein-shifted) is alone not hermitian but only normal. Nevertheless one can still 
compute the heat trace coefficients of $D_w^\dagger D_w$ and then, using the Wick rotation back to the Lorentzian case 
recover the action functional. 

In what follows we assume that we work on a flat compact manifold (torus) so all curvature terms vanish from the spectral action, 
and we are using a physical parametrisation of fields, described next.

\subsection{Spectral action - physical parametrization}
\label{sec:parametrization}
Let us now write explicitly the full spectral action in terms of Yukawa parameters and Higgs field in the case 
of one generation of fermions. Since $\mathbf{\Phi}$ is a quaternion it can be parametrize as 
\begin{equation*}
\mathbf{\Phi}=\begin{pmatrix}
	\phi_1 & \phi_2\\
	-\overline{\phi_2} & \overline{\phi_1}
\end{pmatrix},    
\end{equation*}
where $\phi_1,\phi_2$ are two complex fields. Then 
\begin{equation*}
    \begin{split}
        	\Phi_l=M_l(1+\mathbf{\Phi})=
	\begin{pmatrix}
		\Upsilon_\nu  H_1 & \Upsilon_\nu   H_2 \\
		-\Upsilon_e  \overline{H_2} & \Upsilon_e  \overline{H_1}
	\end{pmatrix}, \\
	\Phi_q=M_q(1+\mathbf{\Phi})=\begin{pmatrix}
		\Upsilon_u   H_1 & \Upsilon_u  H_2 \\
		-\Upsilon_d  \overline{H_2} & \Upsilon_d \overline{H_1}
	\end{pmatrix},
    \end{split}
\end{equation*}
where we introduced the Higgs doublet:
\begin{equation*}
H\equiv\begin{pmatrix}
	H_1\\
	H_2
\end{pmatrix}=\begin{pmatrix}
	1+\phi_1\\ \phi_2
\end{pmatrix}.    
\end{equation*}

\subsection{The spectral action  for the full static SM}

We consider here the Krein-shifted operator for the static simplified Higgs model, computing the coefficients of the spectral action for 
its spatial part only, which is an elliptic operator. For consistency we keep the dimension-related constants in the Gilkey-Seelay-DeWitt 
coefficients the same as for four dimensional case.

The fluctuated Krein-shifted static Dirac operator for the Standard Model splits into the lepton and the quark sector,
after introducing new fields, 
\begin{equation}
\Phi_l:=M_l(1+\mathbf{\Phi}),\qquad \Phi_q:=M_q(1+\mathbf{\Phi}),
\end{equation}
with the lepton part
\begin{equation}
\begin{aligned}
\widetilde{D_L} &=
i \begin{pmatrix}
\1_2 & \\ & - \1_2 \end{pmatrix} \otimes \sigma^j \partial_j 
+
\begin{pmatrix}
& \Phi_l \\ \Phi_l^\dagger &
\end{pmatrix} \otimes\1_2 \\
&+
A_j 
\begin{pmatrix}
\sigma^3-\1_2 & \\ & \1_2 
\end{pmatrix} 
\otimes \sigma^j 
-\begin{pmatrix}
\00_2&\\ & W_j
\end{pmatrix} \otimes \sigma^j,
\end{aligned}
\end{equation}
where we have used the identification $M_4(\CC)\otimes M_2(\CC)\otimes \CC\cong M_4(\CC)\otimes M_2(\CC)$ and therefore omitted 
the third component in the expression above. 

This reproduces the correct hypercharges in the leptonic sector: $0,-2,-1,-1$. We remark that for the left particles, the hypercharges 
are defined as coefficients in terms containing $\widetilde{\sigma}^j$ instead of $\sigma^j$. 

For the quark sector we have:
\begin{equation}
\begin{split}
\widetilde{D_Q} 
=& i \begin{pmatrix} \1_2 & \\ & -\1_2  \end{pmatrix}
\otimes \sigma^j\partial_j  \otimes \1_3 
+ 
\begin{pmatrix} & \Phi_q\\ \Phi_q^\dagger & \end{pmatrix}
\otimes\1_2\otimes \1_3
\\
&+
A_j \begin{pmatrix}
\sigma^3+\frac{1}{3}\1_2 & \\   & -\frac{1}{3}\1_2 \end{pmatrix}
\otimes \sigma^j \otimes \1_3 \\
&
-\begin{pmatrix} \00_2 & \\ & W_j \end{pmatrix}
\otimes \sigma^j\otimes \1_3 
+\begin{pmatrix} \1_2 & \\ & - \1_2  \end{pmatrix}
\otimes \sigma^j \otimes G_j.
\end{split}
\end{equation}
Again, it gives correct hypercharges for quarks: $\frac{4}{3},-\frac{2}{3},\frac{1}{3},\frac{1}{3}$. 
Contributions to the action can be computed separately for the leptonic and the quark sector. The detailed computations are in the Appendix \ref{sec:AA}, here we present the final result in the physical parametrization.

\subsubsection{The full spectral action}
The asymptotic expansion of the spectral action for models on a four dimensional space with a fluctuated Dirac operator $D_\omega$ is 
given by
\begin{equation}
\mathrm{Tr}\left(f \left(\frac{D_\omega}{\Lambda}\right)\right)\sim a_4 f(0)+2\sum\limits_{\substack{0\leq k<4\\ k \text{ even}}}f_{4-k}a_k\frac{\Lambda^{4-k}}{\Gamma\left(\frac{4-k}{2}\right)}+\mathcal{O}(\Lambda^{-1}),
\end{equation}
and reduces simply to
\begin{equation*}
    \mathrm{Tr}\left(f \left(\frac{D_\omega}{\Lambda}\right)\right)\sim a_4 f(0)+2a_0f_4\Lambda^4 +2f_2\Lambda^2 a_2+\mathcal{O}(\Lambda^{-1}),
\end{equation*}
where $a_k$ are the so-called Gilkey-Seeley-DeWitt coefficients and can be computed explicitly -- see e.g. \cite{V_manual,walter} for the detailed discussion. Here $f$ is a sufficiently regular function (see e.g. \cite[Chapter~2]{EIbook}) with $f_k$ being its $k$th moment, and $\Lambda$ is the cut-off parameter.

We start with the model containing only one generation of particles. In this case we get
\begin{equation*}
    a_2=-\frac{\kappa}{4\pi^2} a\int d^4x |H|^2,
\end{equation*}
\begin{equation*}
    a_4=\frac{\kappa}{8\pi^2}\int d^4 x \left[b|H|^4+a \mathrm{Tr}|D_jH|^2 +\frac{20}{9}F^2+\frac{2}{3}\mathrm{Tr}W^2 +\frac{2}{3}\mathrm{Tr}G^2\right],
\end{equation*}
where
\begin{equation*}
    \begin{split}
        a&=|\Upsilon_\nu|^2+|\Upsilon_e|^2+3|\Upsilon_u|^2+3|\Upsilon_d|^2,\\
b&=|\Upsilon_\nu|^4+|\Upsilon_e|^4+3|\Upsilon_u|^4+3|\Upsilon_d|^4,
    \end{split}
\end{equation*}
and $\kappa$ is the normalization of the trace.

In case of three generations we have to change the above coefficients into
\begin{equation*}
    \begin{split}
a&=\mathrm{Tr}(\Upsilon_\nu^\dagger\Upsilon_\nu) +\mathrm{Tr}(\Upsilon_e^\dagger\Upsilon_e)+3\mathrm{Tr}(\Upsilon_u^\dagger\Upsilon_u)+3\mathrm{Tr}(\Upsilon_d^\dagger\Upsilon_d),\\
b&=\mathrm{Tr}(\Upsilon_\nu^\dagger\Upsilon_\nu)^2+\mathrm{Tr}(\Upsilon_e^\dagger\Upsilon_e)^2+3\mathrm{Tr}(\Upsilon_u^\dagger\Upsilon_u)^2+3\mathrm{Tr}(\Upsilon_d^\dagger\Upsilon_d)^2,        
    \end{split}
\end{equation*}
and we have to multiply the terms with field curvatures by a factor of $3$. As a result
\begin{equation*}
a_4=\frac{\kappa}{8\pi^2}\int d^4 x \left[b|H|^4+a \mathrm{Tr}|D_jH|^2 +\frac{20}{3}F^2+2\mathrm{Tr}W^2 +2\mathrm{Tr}G^2\right].    
\end{equation*}
Taking $\kappa=4$ and ignoring the term related to the gravitational constant (i.e. the one $\sim \Lambda^4$) we end up with a model 
with an effective Lagrangian $\mathcal{L}=\mathcal{L}_\text{Higgs} +\mathcal{L}_\text{gauge}$, where
\begin{equation*}
    \mathcal{L}_\text{Higgs}=\frac{bf(0)}{2\pi^2}|H|^4-\frac{2f_2 \Lambda^2 a}{\pi^2}|H|^2 +\frac{af(0)}{2\pi^2}\mathrm{Tr}|D_j H|^2,
\end{equation*}
\begin{equation*}
\mathcal{L}_\text{gauge}=\frac{f(0)}{\pi^2}\left(\frac{10}{3}F^2 + \mathrm{Tr}W^2 +\mathrm{Tr}G^2\right).    
\end{equation*}

The above result is in agreement with the one in \cite[Prop.~11.9]{walter}, for $c=d=e=0$ in the notation used therein. Furthermore, notice also 
that this is consistent (up to an irrelevant global factor) with taking the static part of the Lorentzian Lagrangian for the Standard Model. Indeed, 
we have
\begin{equation*}
    \begin{split}
        &-F_{\mu\nu}F^{\mu\nu}+|D_\mu H|^2- V(H)=-2F_{0j}F^{0j}-F_{jk}F^{jk}+D_0H^\dagger D_0 H -D_jH^\dagger D_jH - V(H)\\
&=-F_{jk}F_{jk}-D_{j}H^\dagger D_j H- V(H)=-\left(F_{jk}F_{jk}+D_{j}H^\dagger D_jH + V(H)\right).
    \end{split}
\end{equation*}
In particular any prediction related to the mass of the Higgs field remains unchanged.

\subsection{Wick rotated model}

Let us start with the full Krein-shifted Dirac operator in the leptonic sector,
\begin{equation*}
   \begin{split}
\widetilde{D}_L&=i\begin{pmatrix}
\1_2\otimes\sigma^\mu & \\
& \1_2\otimes \widetilde{\sigma}^\mu
\end{pmatrix}\partial_\mu+A_\mu \begin{pmatrix}
(\sigma^3-\1_2)\otimes\sigma^\mu &\\
& -\1_2\otimes \widetilde{\sigma}^\mu
\end{pmatrix} 
\\ 
&+ \begin{pmatrix}
\00_4 &\\
& W_\mu \otimes\widetilde{\sigma}^\mu
\end{pmatrix}
+\begin{pmatrix}
& \Phi_l\\
\Phi_l^\dagger
\end{pmatrix}\otimes \1_2.
   \end{split} 
\end{equation*}
Its Lorentzian counterpart is of the form
\begin{equation*}
    \begin{split}
D_L&=
i\begin{pmatrix} & \1_2\otimes \widetilde{\sigma}^\mu \\ \1_2 \otimes \sigma^\mu \end{pmatrix}\partial_\mu 
+ A_\mu 
    \begin{pmatrix} & -\1_2\otimes \widetilde{\sigma}^\mu\\ (\sigma^3-\1_2)\otimes\sigma^\mu \end{pmatrix}  \\
&+\begin{pmatrix} & W_\mu \otimes \widetilde{\sigma}^\mu\\ \00_4 \end{pmatrix}
  + \begin{pmatrix} \Phi_l^\dagger & \\ &\Phi_l \end{pmatrix}\otimes\1_2.        
    \end{split}
\end{equation*}
In what follows we perform a Wick rotation on the level of the algebra of Pauli matrices: $\sigma^j\rightarrow i\sigma^j$ and consequently 
$\widetilde{\sigma}^j \rightarrow -i\sigma^j$, while the $\sigma^0$ remains unchanged. The Wick-rotated Dirac operator in the leptonic sector is then 
of the form
\begin{equation}
\begin{aligned}
D_{L,w}= & i\begin{pmatrix}
&\1_2 \\ \1_2 & \end{pmatrix}\otimes \1_2\partial_0 
+ i\begin{pmatrix} & -i \1_2 \\ i \1_2
\end{pmatrix} \otimes \sigma^j\partial_j  \\
& +A_0\begin{pmatrix} &-\1_2\\ (\sigma^3-\1_2)& \end{pmatrix}\otimes \1_2
+A_j\begin{pmatrix} &i\1_2 \\ i(\sigma^3-\1_2)& \end{pmatrix}\otimes \sigma^j \\
& +\begin{pmatrix} & W_0\\ \00_2 \end{pmatrix}\otimes\1_2 
  - \begin{pmatrix} & iW_j\\ \00_2 & \end{pmatrix}\otimes\sigma^j 
+\begin{pmatrix} \Phi_l^\dagger & \\ &\Phi_l \end{pmatrix}\otimes\1_2.
\end{aligned}
\label{DiracWick}
\end{equation}
For the quark sector we have
\begin{equation}
	\begin{split}
		\widetilde{D}_Q &=i\begin{pmatrix}
			\1_2\otimes\sigma^\mu & \\
			&\1_2\otimes \widetilde{\sigma}^\mu
		\end{pmatrix}\otimes \1_3\partial_\mu +A_\mu \begin{pmatrix}
			\left(\sigma^3+\frac{1}{3}\1_2\right)\otimes\sigma^\mu&\\
			& \frac{1}{3}\1_2\otimes\widetilde{\sigma}^\mu
		\end{pmatrix}\otimes\1_3\\
		&+\begin{pmatrix}
			\1_2\otimes \sigma^\mu & \\
			& \1_2\otimes \widetilde{\sigma}^\mu
		\end{pmatrix}\otimes G_\mu +\begin{pmatrix}
			\00_2 & \\
			& W_\mu\otimes\widetilde{\sigma}^\mu
		\end{pmatrix}\otimes\1_3 +\begin{pmatrix}
			& \Phi_q\\
			\Phi_q^\dagger &
		\end{pmatrix}\otimes \1_2\otimes \1_3.
	\end{split}
\end{equation}
Then
\begin{equation}
	\begin{split}
		D_Q&=i\begin{pmatrix}
			& \1_2\otimes \widetilde{\sigma}^\mu \\
			\1_2\otimes\sigma^\mu &
		\end{pmatrix}\otimes\1_3\partial_\mu +A_\mu \begin{pmatrix}
			& \frac{1}{3}\1_2\otimes\widetilde{\sigma}^\mu \\
			\left(\sigma^3 +\frac{1}{3}\1_2\right)\otimes \sigma^\mu&
		\end{pmatrix}\otimes\1_3\\
		&+\begin{pmatrix}
			&\1_2\otimes \widetilde{\sigma}^\mu \\
			\1_2\otimes \sigma^\mu &
		\end{pmatrix}\otimes G_\mu +\begin{pmatrix}
			& W_\mu \otimes\widetilde{\sigma}^\mu\\
			\00_4 &
		\end{pmatrix}\otimes \1_3 +\begin{pmatrix}
			\Phi_q^\dagger &\\
			& \Phi_q
		\end{pmatrix}\otimes\1_2\otimes\1_3,
	\end{split}
\end{equation}
and after performing the Wick rotation we get
\begin{equation}
\label{DiracWickq}
	\begin{split}
		D_{Q,w}&=i\begin{pmatrix}
			&\1_2\\
			\1_2&
		\end{pmatrix}\otimes\1_2\otimes\1_3\partial_0 + i\begin{pmatrix}
			& -i\1_2\\
			i\1_2&
		\end{pmatrix}\otimes\sigma^j \otimes\1_3\partial_j +\begin{pmatrix}
			\Phi_q^\dagger & \\
			&\Phi_q
		\end{pmatrix}\otimes\1_2\otimes\1_3 \\&
		+A_0\begin{pmatrix}
			&\frac{1}{3}\1_2\\
			\sigma^3+\frac{1}{3}\1_2&
		\end{pmatrix}\otimes\1_2\otimes\1_3+iA_j\begin{pmatrix}
			& -\frac{1}{3}\1_2\\
			\sigma^3 +\frac{1}{3}\1_2 &
		\end{pmatrix}\otimes\sigma^j \otimes\1_3\\
		&+\begin{pmatrix}
			& \1_2\\
			\1_2 &
		\end{pmatrix}\otimes\1_2\otimes G_0 +\begin{pmatrix}
			&-\1_2\\
			\1_2&
		\end{pmatrix}\otimes\sigma^j \otimes iG_j \\
		&+\begin{pmatrix}
			& W_0 \\
			\00_2
		\end{pmatrix}\otimes\1_2\otimes\1_3 + \begin{pmatrix}
			& -iW_j\\
			\00_2 &
		\end{pmatrix}\otimes\sigma^j \otimes \1_3.
	\end{split}
\end{equation}
Again, all further details of the computations are in the appendix \ref{sec:AB}, and in what follows we present only the final expressions for the Wick-rotated model.

\subsubsection{The full spectral action}
We consider now the full model with three generations of particles. In this case, using the above results, we get
\begin{equation}
a_2=\frac{3\kappa}{4\pi^2}a\int d^4x|H|^2,
\end{equation}
and
\begin{equation}
\begin{split}
a_4=\frac{\kappa}{8\pi^2}\int d^4x &\left[b|H|^4-a\mathrm{Tr}|D_\mu|^2+\frac{20}{3}F^2 +2\mathrm{Tr}(W^2)+2\mathrm{Tr}(G^2) \right.\\
&\left.+12\varepsilon^{jkl}F_{jk}F_{0l}-6\varepsilon^{jkl}\mathrm{Tr}(W_{jk}W_{0l})\right],
\end{split}
\end{equation}
where the parameters $a$ and $b$ are as before:
\begin{equation*}
    \begin{split}
a&=\mathrm{Tr}(\Upsilon_\nu^\dagger\Upsilon_\nu) +\mathrm{Tr}(\Upsilon_e^\dagger\Upsilon_e)+3\mathrm{Tr}(\Upsilon_u^\dagger\Upsilon_u)+3\mathrm{Tr}(\Upsilon_d^\dagger\Upsilon_d),\\
b&=\mathrm{Tr}(\Upsilon_\nu^\dagger\Upsilon_\nu)^2+\mathrm{Tr}(\Upsilon_e^\dagger\Upsilon_e)^2+3\mathrm{Tr}(\Upsilon_u^\dagger\Upsilon_u)^2+3\mathrm{Tr}(\Upsilon_d^\dagger\Upsilon_d)^2.        
    \end{split}
\end{equation*}

Notice that by construction these parameters are non-negative. Taking $\kappa=4$ and considering the first terms of the asymptotic expansion (and neglecting the gravitational terms) we end up with the following Lagrangians for gauge fields and the field $H$:
\begin{equation}
\mathcal{L}_{\mathrm{gauge}}=\frac{f(0)}{\pi^2}\left(\frac{10}{3}F^2 +\mathrm{Tr}(W^2)+\mathrm{Tr}(G^2)+6\varepsilon^{jkl}F_{jk}F_{0l}-3\varepsilon^{jkl}\mathrm{Tr}(W_{jk}W_{0l})\right),
\end{equation}
\begin{equation}
\mathcal{L}_H=\frac{bf(0)}{2\pi^2}|H|^4+\frac{6f_2 \Lambda^2}{\pi^2}a|H|^2-\frac{af(0)}{2\pi^2}\mathrm{Tr}|D_\mu H|^2.
\end{equation}
Since the Wick rotation was performed in three spatial directions on the level of Pauli algebra, when going back from the Minkowski signature $(1,3)$ into the Euclidean one we have to change spatial derivatives and coordinates according to $\partial_j\rightarrow -i\partial_j$ and $A_j\rightarrow -i A_j$, respectively, and in order to preserve the spin structure we have to change the Minkowskian structure constants $\varepsilon_{M}^{jkl}$ into their Euclidean counterparts: $\varepsilon^{jkl}_E=-i\varepsilon^{jkl}_M$. As a result
\begin{equation*}
    \begin{split}
        -F_{\mu\nu}^M F^{\mu\nu}_M&=-2F_{0j}^MF^{0j}_M -F^M_{jk}F_M^{jk} =2F_{0j}^MF_{0j}^M -F_{jk}^MF_{jk}^M \\
&\rightarrow -2F_{0j}^E F_{0j}^E -F_{jk}^EF_{jk}^E=-F_{\mu\nu}^EF_{\mu\nu}^E,
    \end{split}
\end{equation*}
and 
\begin{equation}
    \begin{split}
&(D_\mu H^\dagger_M )(D^\mu H_M)=(D_0 H^\dagger_M)( D_0 H_M) -(D_j H^\dagger_M )(D_j H_M)\\
&\rightarrow (D_0 H^\dagger_E) (D_0 H_E) +(D_j H^\dagger_E )(D_j H_E) =(D_\mu H^\dagger_E)( D_\mu H_E),        
    \end{split}
\end{equation}
so that for these terms we have
\begin{equation*}
-F_M^2 +|D_\mu H_M|^2-V(H_M) \rightarrow -\left(F_E^2 -|D_\mu H_E|^2+V(H_E)\right)    
\end{equation*}
in a complete agreement with our result. The global minus sign (together with the additional $-i$ factor from the measure) is related to the definition of an Euclidean action: $iS_M=-S_E$. Next, let us consider the remaining term:
\begin{equation*}
\varepsilon^{\mu\nu\rho\sigma}_M F^{M}_{\mu\nu} F^M_{\rho\sigma}=4\varepsilon^{jkl}_MF_{0j}^MF_{kl}^M\rightarrow -4 \varepsilon^{jkl}_E F^{E}_{0j}F^E_{jk}.    
\end{equation*}
Taking into account the additional global sign from the identification of $iS_M$ with $-S_E$, we finally see that the Lorentzian counterpart of $6\varepsilon^{jkl}F_{jk}F_{0l}-3\varepsilon^{jkl}\mathrm{Tr}\left(W_{jk}W_{0l}\right)$
is 
\begin{equation*}
\frac{3}{2}\varepsilon^{\mu\nu\rho\sigma}F_{\mu\nu}F_{\rho\sigma}-\frac{3}{4}\varepsilon^{\mu\nu\rho\sigma}\mathrm{Tr}\left(W_{\mu\nu}W_{\rho\sigma}\right).    
\end{equation*}
Therefore, the spectral action for this model contains terms which can be interpreted as the so-called $\theta$-terms in the electroweak sector \cite{MS98, PP2014, KM2021}. We remark that from the above derivation of the spectral action not only the presence of such terms is deduced but also the numerical value of the electroweak vacuum angle is fixed by the model. The presence of such terms was linked with the CP-violation \cite{PP2014}, especially for the discussion of the baryogenesis process. In contrast to the usual considerations in the physical formulation of the Standard Model no CP-violating $\theta$-term in the QCD sector is present here. It will be interesting to see what are the physical limitations, e.g. on the energy scales on which such model is valid, from the perspective of the presence of the electroweak $\theta$-terms. The CP-violation was present in this model also on the level of algebra by the lack of the $\mathcal{J}$-symmetry \cite{BS20}.  

We remark that the explicit form of the potential $V(H)$ differs from the one in the standard derivation \cite{walter}, where the coefficient in the quadratic term $|H|^2$ contained $-2f_2
$ instead of $6f_2$, which we have in the present model. In the usual formulation $f$ is assumed to be, besides the others, a non-negative on the positive real half-line, so then $f_2$ is also non-negative therein. If we would not allowed for any relaxation of this principle, our model will not predict the Higgs mechanism, or in other words the model could be valid only in a sector with the Higgs potential of the form $|H|^4+b_1|H|^2+b_2$ with positive $b_1,b_2$, i.e. the Higgs potential will not possess a non-trivial minimum. On the other hand, having the possibility of using function $f$ which has negative second moment gives rise to an effective action for the Standard Model with the Higgs mechanism implemented in a completely similar manner as in the usual product-like almost-commutative geometry \cite{walter}. Since all the derivations were made on the algebraic level we could, by linearity, extend the applicability of the usual methods into the case with $f$ being a difference of two positive functions. However, the discussion of the analytical aspects is required in order to establish the range of validity of these computational methods -- see \cite{EIbook} for some further discussion of these aspects which are beyond the scope of this paper.

Allowing for the negative value of $f_2$ there is no further difference in the numerical value of the Higgs mass, which can be computed from the derived Lagrangian using the standard tools based on the renormalization group equation \cite{CCM07, walter}. This is because the difference in the numerical value of $f_2$ in the coefficient for the $|H|^2$ term does not affect any relation between the mass of the $W$ boson, the Higgs vacuum expectation value $v$ and the coupling constant $g_2$ for the $W$ boson field.  

\section{Conclusions and outlook}

The presented noncommutative geometric model describing the particle interaction appears to be the closest to the observed
Standard Model. The derived bosonic spectral action gives all correct terms and, in addition, new, topological $\theta$-terms.
While the latter has no significance for the dynamics of the model, it may play a role for the quantum effects for the electroweak
sector. These terms are, in principle, not excluded and have been discussed in literature \cite{MS98, PP2014, KM2021}. The result 
signifies also that computing the spectral action for the Wick-rotated Lorentzian Dirac operator is important. Possibly, the next step 
shall be to compute the genuine Lorentzian spectral action using the tools that are at present available for the Laplace-type
operators \cite{Wr21}. Furthermore, possible relations of non-product geometries with bundle-like structures over 
noncommutative manifolds \cite{BoDu14} as well to the inclusion of gravity for this non-product geometry (see \cite{BS21} for
a link between nonproduct geometries and gravity) shall also be explored and examined.  Finally, it shall be interesting to 
see possible extensions of the model, both in the direction of scalar conformal modifications that can help to fix the Higgs mass 
as well as extensions of the Pati-Salam type \cite{PS1,PS2}.

\appendix
\section{The static spectral action}
\label{sec:AA}
\subsection{Leptonic sector}
In the leptonic sector we have
\begin{equation}
	\widetilde{D_L}^2 =
	- (\1_4 \otimes \1_2) \Delta  
	-a^j \partial_j  - b,  
\end{equation}
where,
\begin{equation}
a^j=-2i\left( 
	A_j \begin{pmatrix}	\sigma^3-\1_2 & \\ & -\1_2 \end{pmatrix} 
	+ \begin{pmatrix}	\00_2&\\ & W_j \end{pmatrix}  \right) \otimes \1_2,
\end{equation}
\begin{equation}
	\begin{split}
		b = &\phantom{+}-
		\begin{pmatrix} \Phi_l {\Phi_l}^\dagger &\\ &{ \Phi_l}^\dagger \Phi_l \end{pmatrix} \otimes \1_2 
		- A_j A_k \begin{pmatrix} 2 (\1_2 -\sigma^3) &\\ & \1_2  \end{pmatrix} \otimes \sigma^j \sigma^k  \\
		& - \begin{pmatrix} \00_2 & \\ &  W_j W_k  \end{pmatrix} \otimes \sigma^j \sigma^k + \begin{pmatrix} & \Phi_l  W_j \\ W_j {\Phi_l}^\dagger \end{pmatrix} \otimes \sigma^j 
		+2 \begin{pmatrix}	\00_2&\\ & W_j \end{pmatrix}   A_j \otimes \1_2 \\
		& - i \begin{pmatrix} & \partial_j\Phi_l \\ - \partial_j\Phi_l^\dagger \end{pmatrix} \otimes \sigma^j 
		- i \begin{pmatrix} \00_2 &\\ & \partial_j W_k \end{pmatrix} \otimes \sigma^j\sigma^k  \\
		& - i \begin{pmatrix}	\sigma^3-\1_2 & \\ & -\1_2 \end{pmatrix} \partial_j A_k  \otimes \sigma^j\sigma^k  
		- \begin{pmatrix} & \sigma^3 \Phi_l  \\ {\Phi_l}^\dagger \sigma^3 \end{pmatrix} A_j \otimes \sigma^j.
	\end{split}
\end{equation}
As a result, following the notation of \cite{V_manual}, we have $ \omega_j=\frac{1}{2}a^j$, so that
\begin{equation}
\begin{split}
	\Omega_{ij}&=\partial_i\omega_j-\partial_j\omega_i+\omega_i\omega_j-\omega_j\omega_i\\
	&= - i F_{ij} \begin{pmatrix}	\sigma^3-\1_2 & \\ & -\1_2 \end{pmatrix} \otimes\1_2 - i \begin{pmatrix} \00_2 &\\ &W_{ij} \end{pmatrix}\otimes\1_2,
\end{split}
\end{equation}
with
\begin{equation}
	F_{ij}=\partial_i A_j-\partial_j A_i, \qquad W_{ij}=\partial_iW_j-\partial_j W_i-i[W_i,W_j].
\end{equation}

Next we compute,
\begin{equation}
	\begin{split}
		E  &=b-\partial_j\omega_j-\omega_j\omega_j\\
		&=- \begin{pmatrix}
			\Phi_l\Phi_l^\dagger & \\
			& \Phi_l^\dagger \Phi_l
		\end{pmatrix}\otimes\1_2 - i \begin{pmatrix}
			& \partial_j\Phi_l \\
			-\partial_j \Phi_l^\dagger &
		\end{pmatrix}\otimes \sigma^j \\
		&+\frac{1}{2}\begin{pmatrix}
			\00_2 & \\
			& W_{jk}
		\end{pmatrix}\otimes \varepsilon^{jkl}\sigma^l +\frac{1}{2}F_{jk} \begin{pmatrix}
			\sigma^3-\1_2 &\\
			& -\1_2
		\end{pmatrix}\otimes \varepsilon^{jkl}\sigma^l \\
		&-A_j\begin{pmatrix}
			& \sigma^3\Phi_l\\
			\Phi_l^\dagger \sigma^3 &
		\end{pmatrix}\otimes \sigma^j +\begin{pmatrix}
			&\Phi_l W_j \\
			W_j\Phi_l^\dagger
		\end{pmatrix}\otimes \sigma^j.
	\end{split}
\end{equation}
We get then
\begin{equation}
	\hbox{Tr}(E)=-4\kappa \mathrm{Tr}(\Phi_l^\dagger \Phi_l) , 
\end{equation}
and furthermore,
\begin{equation}
	\hbox{Tr}(\Omega_{ij}\Omega_{ij})=-2\kappa \left(6F^2 +\mathrm{Tr}(W^2)\right),
\end{equation}
where $\kappa$ is the normalization of the trace (i.e. everything within the bracket is computed for the unnormalized trace) and $W^2=W_{jk}W_{jk}$. Moreover,
\begin{equation}
	\begin{split}
		\kappa^{-1}\hbox{Tr}(E^2)&=4\mathrm{Tr}(\Phi_l^\dagger\Phi_l)^2 +4\mathrm{Tr}[(\partial_j\Phi_l^\dagger)(\partial_j\Phi_l)]+4A^2\mathrm{Tr}(\Phi_l^\dagger\Phi_l)+4\mathrm{Tr}(W_jW_j\Phi_l^\dagger \Phi_l)\\
		&+4iA_j\mathrm{Tr}\left[((\partial_j\Phi_l)\Phi_l^\dagger -\Phi_l(\partial_j\Phi_l^\dagger) )\sigma^3\right]-4i\mathrm{Tr}\left[(\Phi_l^\dagger (\partial_j\Phi_l)-(\partial_j\Phi_l^\dagger)\Phi_l )W_j\right]\\
		&-8A_j\mathrm{Tr}\left(\Phi_l^\dagger \sigma^3\Phi_lW_j\right)
		+6F^2 + \mathrm{Tr}(W^2).
	\end{split}
\end{equation}
As a result, in the leptonic sector we have
\begin{equation}
	\begin{split}
		a_2 &= \frac{1}{(4\pi)^2} \int d^4x \, \mathrm{Tr} E = 
		- \frac{\kappa}{4\pi^2}\int d^4x \, \mathrm{Tr}(\Phi_l^\dagger \Phi_l), \\
		a_4 & = \frac{1}{16\pi^2} \frac{1}{12} \int d^4x \, \left( 6 \mathrm{Tr} (E^2) + \hbox{Tr}(\Omega_{ij}\Omega_{ij}) \right)   \\
		& = \frac{\kappa}{48\pi^2}\int d^4x \left[ 6 \left( \mathrm{Tr}(\Phi_l^\dagger\Phi_l)^2  
		+\mathrm{Tr}[(\partial_j\Phi_l^\dagger)(\partial_j\Phi_l)]+A^2\mathrm{Tr}(\Phi_l^\dagger\Phi_l)\right.\right.\\
		&\left.\left. \qquad  +\mathrm{Tr}(W_jW_j\Phi_l^\dagger \Phi_l)
		+iA_j\mathrm{Tr}\left[((\partial_j\Phi_l)\Phi_l^\dagger -\Phi_l(\partial_j\Phi_l^\dagger) )\sigma^3\right]
		\right.\right.\\ & \left.\left. 
		\qquad -i\mathrm{Tr}\left[(\Phi_l^\dagger (\partial_j\Phi_l)-(\partial_j\Phi_l^\dagger)\Phi_l )W_j\right] 
		-2A_j\mathrm{Tr}\left(\Phi_l^\dagger \sigma^3\Phi_lW_j\right) \right) \right. \\
		& \left. \qquad +6F^2+\mathrm{Tr}(W^2)\right].
	\end{split}
\end{equation}
Using the parametrization from the section \ref{sec:parametrization} in this sector we then have
\begin{equation*}
\Phi_l^\dagger \Phi_l =
\begin{pmatrix}
	|\Upsilon_\nu|^2 |H_1|^2   + |\Upsilon_e|^2  |H_2|^2
	& |\Upsilon_\nu|^2  \overline{H_1} H_2 - |\Upsilon_e|^2 H_2 \overline{H_1}\\
	|\Upsilon_\nu|^2  \overline{H_2} H_1- |\Upsilon_e|^2 H_1\overline{H_2} 
	&  |\Upsilon_\nu|^2  |H_2|^2 + |\Upsilon_e|^2 |H_1|^2 
\end{pmatrix},    
\end{equation*}
and as a result
\begin{equation*}
a_2=-\frac{\kappa}{4\pi^2}(|\Upsilon_e|^2+|\Upsilon_\nu|^2)\int d^4x \,  |H|^2.    
\end{equation*}
Furthermore we have
\begin{equation*}
\begin{split}
	\mathrm{Tr}(\Phi_l^\dagger \Phi_l)^2 &= 
	(|\Upsilon_\nu|^4+|\Upsilon_e|^4)|H|^4,
\end{split}    
\end{equation*}
\begin{equation*}
    \mathrm{Tr}\left[(\partial_j\Phi_l^\dagger)(\partial_j\Phi_l)\right]=
(|\Upsilon_\nu|^2+|\Upsilon_e|^2) \, |\partial_j H|^2,
\end{equation*}
\begin{equation*}
(A_j A_j) \mathrm{Tr}(\Phi_l^\dagger\Phi_l)= (A_j A_j) (|\Upsilon_\nu|^2+|\Upsilon_e|^2)\, |H|^2.    
\end{equation*}

Next, we decompose the $W$ field in terms of Pauli matrices,
\begin{equation*}
W_j=W_{j,1}\sigma^1+W_{j,2}\sigma^2+W_{j,3}\sigma^3,    
\end{equation*}
so that
\begin{equation*}
\mathrm{Tr}(W_jW_j\Phi_l^\dagger \Phi_l)=  
(W_jW_j) (|\Upsilon_\nu|^2+|\Upsilon_e|^2)|H|^2.    
\end{equation*}

Next, we compute
\begin{equation*}
iA_j\mathrm{Tr}\left[((\partial_j\Phi_l)\Phi_l^\dagger -\Phi_l(\partial_j\Phi_l^\dagger) )\sigma^3\right]=
iA_j  (|\Upsilon_\nu|^2+|\Upsilon_e|^2)(H^\dagger \partial_j H -\partial_j H^\dagger H),    
\end{equation*}
\begin{equation*}
\begin{split}
	-i\mathrm{Tr} & \left[(\Phi_l^\dagger (\partial_j\Phi_l)-(\partial_j\Phi_l^\dagger)\Phi_l )W_j \right] = \\
	&= -i  (|\Upsilon_\nu|^2+|\Upsilon_e|^2) \left[W_{j,3}\left(\overline{H_1} \partial_j H_1 
	- H_1\partial_j \overline{H_1} - \overline{H_2}\partial_j H_2 +H_2\partial_j\overline{H_2}\right)\right. \\
	&\left. \qquad \qquad+ (W_{j,1}-iW_{j,2})(\overline{H_2}\partial_jH_1-H_1\partial_j\overline{H_2})\right.\\
	&\left. \qquad \qquad + (W_{j,1}+iW_{j,2})(\overline{H_1} \partial_j H_2 -H_2\partial_j \overline{H_1})
	\right],
\end{split}    
\end{equation*}
\begin{equation*}
    \begin{split}
        	-2A_j\mathrm{Tr}\left[\Phi_l^\dagger \sigma^3 \Phi_l W_j\right] = & -2A_j(|\Upsilon_\nu|^2+|\Upsilon_e|^2) \left[(W_{j,1}-iW_{j,2})H_1\overline{H_2} \right. \\
	&\left. +(W_{j,1}+iW_{j,2})\overline{H_1} H_2  + W_{j,3}(|H_1|^2-|H_2|^2)\right].
    \end{split}
\end{equation*}

Let us now verify whether these terms can be written using the covariant derivative of the Higgs doublet,
\begin{equation*}
    D_j H=\partial_j H + i W_{j} H - iA_j H.
\end{equation*}
We check,
\begin{equation*}
    \begin{split}
	\mathrm{Tr}|D_jH|^2 &=\mathrm{Tr} \bigl[ |\partial_j H|^2 + i(\partial_j H^\dagger W_j H-H^\dagger W_j\partial_j H)  \\
	&+i A_j(H^\dagger \partial_j H-\partial_j H^\dagger H) - 2A_j H^\dagger W_j H + W_jW_j |H|^2 + A^2|H|^2 \bigr].        
    \end{split}
\end{equation*}
The only terms that are potentially different that the ones in the coefficient $a_4$ are
\begin{equation*}
2A_j\mathrm{Tr}(H^\dagger W_j H), \qquad i\mathrm{Tr}(\partial_j H^\dagger W_j H -H^\dagger W_j \partial_j H),    
\end{equation*}
but simple calculation shows that
\begin{equation*}
    \begin{split}
        	A_j\mathrm{Tr}(H^\dagger W_j H)= &	A_j\bigl[ W_{j,1}(\overline{H_1} H_2-\overline{H_2} H_1) \\
	& +i W_{j,2}(\overline{H_1} H_2 - \overline{H_2} H_1)
	+ W_{j,3}(|H_1|^2-|H_2|^2) \bigr],
    \end{split}
\end{equation*}
and
\begin{equation*}
    \begin{split}
    \mathrm{Tr}(\partial_j H^\dagger W_j H -H^\dagger W_j \partial_j H)&=W_{j,1}(\partial_j \overline{H_1} H_2+\partial_j \overline{H_2} H_1-\overline{H_1} \partial_j H_2-\overline{H_2}\partial_j H_1)\\
	&+W_{j,2}(\partial_j \overline{H_1}H_2-\partial_j \overline{H_2} H_1-\overline{H_1} \partial_j H_2 +\overline{H_2} \partial_j H_1)\\
	&+W_{j,3}(\partial_j \overline{H_1} H_1-\partial_j \overline{H_2} H_2-\overline{H_1} \partial_j H_1+\overline{H_2} \partial_j H_2)    
    \end{split}
\end{equation*}
in a complete agreement with $a_4$. 

Therefore,
\begin{equation}
	a_4 = \frac{\kappa}{8\pi^2}\int \!d^4 x\,  \left[ (|\Upsilon_\nu|^4 \!+\! |\Upsilon_e|^4)|H|^4 \!+\! (|\Upsilon_\nu|^2 \!+\!|\Upsilon_e|^2)\mathrm{Tr}|D_j H|^2  \!+\! F^2   \!+\! \frac{1}{6}\mathrm{Tr}W^2 \right].
\end{equation}
\subsection{Quark sector}

In this sector we have
\begin{equation}
	\widetilde{D}_Q^2=-(\1_4\otimes\1_2\otimes \1_3)\Delta-a^j\partial_j -b,
\end{equation}
where
\begin{equation*}
	a^j=-2i \left[A_j\begin{pmatrix}
		\sigma^3+\frac{1}{3}\1_2 &\\
		& \frac{1}{3}\1_2
	\end{pmatrix}\otimes\1_2\otimes\1_3 + \begin{pmatrix}
		\00_2&\\ &W_j
	\end{pmatrix}\otimes\1_2\otimes\1_3+\1_4\otimes\1_2\otimes G_j \right],
\end{equation*}
\begin{equation*}
	\begin{split}
		b&=-\begin{pmatrix}
			\Phi_q\Phi_q^\dagger & \\
			& \Phi_q^\dagger \Phi_q
		\end{pmatrix}\otimes\1_2\otimes\1_3-A_jA_k\begin{pmatrix}
			\frac{8}{9}\1_2& \\
			& -\frac{1}{9}\1_2
		\end{pmatrix}\otimes \sigma^j\sigma^k\otimes\1_3\\
		&-\begin{pmatrix}
			\00_2 &\\
			& W_jW_k
		\end{pmatrix}\otimes \sigma^j\sigma^k \otimes\1_3 -\1_4 \otimes\sigma^j\sigma^k\otimes G_jG_k\\
		&-i\begin{pmatrix}
			& \partial_j \Phi_q\\
			-\partial_j\Phi_q^\dagger &
		\end{pmatrix}\otimes \sigma^j\otimes\1_3 - i\begin{pmatrix}
			\00_2&\\
			& \partial_j W_k
		\end{pmatrix}\otimes \sigma^j\sigma^k \otimes\1_3\\
		&-i 1_4\otimes \sigma^j\sigma^k \otimes \partial_j G_k -A_j\begin{pmatrix}
			& \sigma^3 \Phi_q\\
			\Phi_q^\dagger\sigma^3 &
		\end{pmatrix}\otimes \sigma^j\otimes\1_3\\
		&+\begin{pmatrix}
			& \Phi_q W_j\\
			W_j\Phi_q^\dagger&
		\end{pmatrix}\otimes\sigma^j\otimes\1_3-i(\partial_j A_k)\begin{pmatrix}
			\sigma^3+\frac{1}{2}\1_2& \\
			&\frac{1}{3}\1_2
		\end{pmatrix}\otimes \sigma^j\sigma^k\otimes \1_3\\
		&-2A_j\begin{pmatrix}
			\sigma^3+\frac{1}{2}\1_2 & \\
			& \frac{1}{3}\1_2
		\end{pmatrix}\otimes \1_2 \otimes G_j - 2\begin{pmatrix}
			\00_2& \\
			& W_{j}
		\end{pmatrix}\otimes \1_2\otimes G_j\\
		&-\frac{2}{3}A_jA_k\begin{pmatrix}
			\sigma^3 &\\
			& \00_2
		\end{pmatrix}\otimes \sigma^j\sigma^k\otimes\1_3 -\frac{2}{3}\begin{pmatrix}
			\00_2& \\ &  A_jW_j
		\end{pmatrix}\otimes \1_2\otimes \1_3.
	\end{split}
\end{equation*}
Therefore,
\begin{equation}
	\begin{split}
		E&=b-\partial_j\omega_j-\omega_j\omega_j\\
		&=-\begin{pmatrix}
			\Phi_q\Phi_q^\dagger & \\
			& \Phi_q^\dagger \Phi_q
		\end{pmatrix}\otimes\1_2\otimes\1_3-i\begin{pmatrix}
			& \partial_j\Phi_q \\
			-\partial_j\Phi_q^\dagger
		\end{pmatrix}\otimes \sigma^j\otimes\1_3\\
		&+\frac{1}{2}F_{jk}\begin{pmatrix}
			\sigma^3&\\
			& \00_2
		\end{pmatrix}\otimes \varepsilon^{jkl}\sigma^l \otimes\1_3 +\frac{1}{2}\begin{pmatrix}
			\00_2 &\\
			& W_{jk}
		\end{pmatrix}\otimes \varepsilon^{jkl}\sigma^l \otimes\1_3+\frac{1}{2}\1_4\otimes \varepsilon^{jkl}\sigma^l\otimes G_{jk}\\
		&-A_j\begin{pmatrix}
			&\sigma^3\Phi_q\\
			\Phi_q^\dagger \sigma^3&
		\end{pmatrix}\otimes\sigma^j\otimes \1_3 +\begin{pmatrix}
			&\Phi_qW_j\\
			W_j\Phi_q^\dagger&
		\end{pmatrix}\otimes \sigma^j\otimes \1_3\\
		&+\frac{1}{6}F_{jk}\1_4\otimes \varepsilon^{jkl}\sigma^l\otimes \1_3,
	\end{split}
\end{equation}
where again
\begin{equation}
	G_{ij}=\partial_i G_j-\partial_j G_i -i[G_i,G_j].
\end{equation}
As a result,
\begin{equation}
	\kappa^{-1}\hbox{Tr}(E)=-12 \mathrm{Tr}(\Phi_q^\dagger \Phi_q).
\end{equation}
and
\begin{equation}
	\kappa^{-1}\hbox{Tr}(\Omega_{ij}\Omega_{ij})=-2\left(\frac{22}{3}F^2+3\mathrm{Tr}(W^2)+4\mathrm{Tr}(G^2)\right),
\end{equation}
where we use short notation $G^2 = G_{ij} G_{ij}$, and the full second contribution reads,  
\begin{equation}
	\begin{split}
		\kappa^{-1}\hbox{Tr}(E^2) = 
		& 12\mathrm{Tr}(\Phi_q^\dagger\Phi_q)^2 
		+ 12\mathrm{Tr}[(\partial_j\Phi_q^\dagger)(\partial_j\Phi_q)] 
		+12A^2\mathrm{Tr}(\Phi_q^\dagger\Phi_q) \\ 
		& +12\mathrm{Tr}(W_jW_j\Phi_q^\dagger \Phi_q) 
		+12iA_j\mathrm{Tr}\left[((\partial_j\Phi_q)\Phi_q^\dagger 
		- \Phi_q(\partial_j\Phi_q^\dagger) )\sigma^3\right] \\
		& - 12i\mathrm{Tr}\left[(\Phi_q^\dagger (\partial_j\Phi_q)
		-(\partial_j\Phi_q^\dagger)\Phi_q )W_j\right]
		-24A_j\mathrm{Tr}\left(\Phi_q^\dagger \sigma^3\Phi_qW_j\right) \\
		& + \frac{22}{3}F^2 + 3\mathrm{Tr}(W^2)+4\mathrm{Tr}(G^2).
	\end{split}
\end{equation}
As a result, in the quark sector we have
\begin{equation}
	a_2=-\frac{\kappa}{4\pi^2}\int d^4x\, 3 \mathrm{Tr}(\Phi_q^\dagger \Phi_q),
\end{equation}
\begin{equation}
	\begin{split}
		a_4 = & \frac{\kappa}{48\pi^2}\int d^4x 
		\left[18\left( \mathrm{Tr}(\Phi_q^\dagger\Phi_q)^2 
		+\mathrm{Tr}[(\partial_j\Phi_q^\dagger)(\partial_j\Phi_q)]
		+A^2\mathrm{Tr}(\Phi_q^\dagger\Phi_q) 
		\right.\right.\\
		&\left.\left.
		+\mathrm{Tr}(W_jW_j\Phi_q^\dagger \Phi_q)
		+ iA_j\mathrm{Tr}\left[((\partial_j\Phi_q)\Phi_q^\dagger 
		-\Phi_q(\partial_j\Phi_q^\dagger) )\sigma^3\right]
		\right.\right.\\ &\left.\left.
		-i\mathrm{Tr}\left[(\Phi_q^\dagger (\partial_j\Phi_q)-(\partial_j\Phi_q^\dagger)\Phi_q )W_j\right] 
		-2A_j\mathrm{Tr}\left(\Phi_q^\dagger \sigma^3\Phi_qW_j\right) \right) 
		.\right.\\ &\left. 
		+ \frac{22}{3}F^2+3\mathrm{Tr}(W^2)+4\mathrm{Tr}(G^2)\right].
	\end{split}
\end{equation}
In a completely similar manner as for the leptonic sector we derive:
\begin{equation}
	a_2=-\frac{\kappa}{4\pi^2}(3|\Upsilon_u|^2+3|\Upsilon_d|^2)\int d^4x |H|^2
\end{equation}
and
\begin{equation}
	\begin{split}
		a_4=\frac{\kappa}{8\pi^2}\int d^4 x\,  &\left[ \left(3|\Upsilon_u|^4 + 3|\Upsilon_d|^4\right)|H|^4+ (3|\Upsilon_u|^2+3|\Upsilon_d|^2)\mathrm{Tr}|D_j H|^2 \right.\\
		&\left.+\frac{11}{9}F^2 +\frac{1}{2}\mathrm{Tr}W^2 +\frac{2}{3}\mathrm{Tr}G^2 \right].
	\end{split}
\end{equation}

\section{The Wick rotated model}
\label{sec:AB}
\subsection{Leptonic sector}
Starting with the Wick rotated Dirac operator \eqref{DiracWick} we get
\begin{equation*}
    \begin{split}
        D_{L,w}^\dagger D_{L,w}&=-(\1_4\otimes\1_2)\Delta_E +2i \left[A_0\begin{pmatrix}
		\sigma^3-\1_2&\\
		&-\1_2
	\end{pmatrix}\otimes \1_2+\begin{pmatrix}
		\00_2 & \\
		& W_0
	\end{pmatrix}\otimes \1_2+\begin{pmatrix}
		& \Phi_l\\
		\Phi_l^\dagger&
	\end{pmatrix}\otimes\1_2\right]\partial_0 \\
	&+2i \left[A_j\begin{pmatrix}
		\sigma^3-\1_2&\\
		&-\1_2
	\end{pmatrix}\otimes \1_2+\begin{pmatrix}
		\00_2 & \\
		& W_j
	\end{pmatrix}\otimes \1_2+
	\begin{pmatrix}
		& -i\Phi_l\\
		i\Phi_l^\dagger
	\end{pmatrix}\otimes \sigma^j
	\right]\partial_j\\
	&+i(\partial_0 A_0)\begin{pmatrix}
		\sigma^3-\1_2&\\
		& -\1_2
	\end{pmatrix}\otimes\1_2 +i(\partial_j A_k)\begin{pmatrix}
		\sigma^3-\1_2 & \\
		& -\1_2
	\end{pmatrix}\otimes\sigma^j\sigma^k\\
	&+A_0^2\begin{pmatrix}
		2(\1_2-\sigma^3) &\\
		& \1_2
	\end{pmatrix}\otimes\1_2+A_jA_k\begin{pmatrix}
		2(\1_2-\sigma^3) &\\
		& \1_2
	\end{pmatrix}\otimes\sigma^j\sigma^k\\
	&+\begin{pmatrix}
		\00_2 & \\
		&W_0^2 + i\partial_0W_0
	\end{pmatrix}\otimes \1_2+\begin{pmatrix}
		\00_2&\\
		& W_jW_k+i\partial_jW_k
	\end{pmatrix}\otimes\sigma^j\sigma^k\\
	&-F_{0j}\begin{pmatrix}
		\sigma^3-\1_2 & \\
		& \1_2
	\end{pmatrix}\otimes\sigma^j+\begin{pmatrix}
		\00_2 & \\
		& W_{0j}
	\end{pmatrix}\otimes \sigma^j-2\begin{pmatrix}
		\00_2 & \\
		& A_0W_0+A_jW_j
	\end{pmatrix}\otimes\1_2\\
	&+i\begin{pmatrix}
		& \partial_0\Phi_l \\
		\partial_0\Phi_l^\dagger &
	\end{pmatrix}\otimes\1_2 +i\begin{pmatrix}
		& -i\partial_j\Phi_l\\
		i\partial_j \Phi_l^\dagger
	\end{pmatrix}\otimes \sigma^j +\begin{pmatrix}
		\Phi_l\Phi_l^\dagger & \\
		&\Phi_l^\dagger \Phi_l
	\end{pmatrix}\otimes\1_2\\
	&+A_0\begin{pmatrix}
		& (\sigma^3-2\cdot \1_2)\Phi_l\\
		\Phi_l^\dagger(\sigma^3-2\cdot\1_2)&
	\end{pmatrix}\otimes\1_2+\begin{pmatrix}
		& \Phi_l W_0\\
		W_0\Phi_l^\dagger
	\end{pmatrix}\otimes\1_2\\
	&+A_j\begin{pmatrix}
		&-i(\sigma^3-2\cdot \1_2)\Phi_l\\
		i\Phi_l^\dagger(\sigma^3-2\cdot \1_2)&
	\end{pmatrix}\otimes\sigma^j+\begin{pmatrix}
		& -i\Phi_l W_j \\
		i W_j \Phi_l^\dagger
	\end{pmatrix} \otimes \sigma^j.
    \end{split}
\end{equation*}
Writing $D_{L,w}^\dagger D_{L,w}$ in the canonical form $-(\1_4\otimes\1_2)\Delta_E -2\omega_\mu\partial_\mu -b$ (with the Euclidean summation) we get 
\begin{equation}
	\begin{split}
		E&=\frac{1}{2}F_{jk}\varepsilon^{jkl}\begin{pmatrix}
			\sigma^3-\1_2 & \\
			& -\1_2
		\end{pmatrix}\otimes\sigma^l + F_{0j}\begin{pmatrix}
			\sigma^3-\1_2 &\\
			& \1_2
		\end{pmatrix}\otimes\sigma^j \\
		&+\frac{1}{2}\varepsilon^{jkl}\begin{pmatrix}
			\00_4 & \\
			& W_{jk}
		\end{pmatrix}\otimes \sigma^l -\begin{pmatrix}
			\00_4 & \\
			& W_{0j}
		\end{pmatrix}\otimes\sigma^j +3\begin{pmatrix}
			\Phi_l\Phi_l^\dagger & \\
			&\Phi_l^\dagger \Phi_l
		\end{pmatrix}\otimes\1_2.
	\end{split}
\end{equation}
Its trace is therefore 
\begin{equation}
	{\hbox{ Tr}}(E)=12\kappa \mathrm{Tr}(\Phi_l^\dagger \Phi_l^\dagger).
\end{equation}
Furthermore, we have
\begin{equation}
	\kappa^{-1}\hbox{Tr}(E^2)=6F^2+\mathrm{Tr}(W^2) +36\mathrm{Tr}(\Phi_l^\dagger \Phi_l)^2 +4 \varepsilon^{jkl}F_{jk}F_{0l}-2\varepsilon^{jkl}\mathrm{Tr}(W_{jk}W_{0l}), 
\end{equation}
where now $F^2= F_{\mu\nu}F_{\mu\nu}=F_{jk}F_{jk}+2F_{0j}F_{0j}$ and similarly for $W^2$.

Next, we have
\begin{equation}
	\begin{split}
		\Omega_{0j}&=-i F_{0j}\begin{pmatrix}
			\sigma^3-\1_2 & \\
			&-\1_2
		\end{pmatrix}\otimes\1_2 - i \begin{pmatrix}
			\00_2 & \\
			& W_{0j}
		\end{pmatrix}\otimes \1_2 +iA_0 \begin{pmatrix}
			& \sigma^3\Phi_l \\
			\Phi_l^\dagger \sigma^3
		\end{pmatrix}\otimes \sigma^j\\
		&-i\begin{pmatrix}
			& \Phi_lW_0\\
			W_0\Phi_l^\dagger
		\end{pmatrix}\otimes\sigma^j -A_j\begin{pmatrix}
			& -\sigma^3\Phi_l \\
			\Phi_l^\dagger \sigma^3
		\end{pmatrix}\otimes\1_2 +\begin{pmatrix}
			& -\Phi_lW_j\\
			W_j\Phi_l^\dagger
		\end{pmatrix}\otimes\1_2\\
		&-2i\begin{pmatrix}
			\Phi_l\Phi_l^\dagger &\\
			&-\Phi_l^\dagger\Phi_l
		\end{pmatrix}\otimes\sigma^j +i \begin{pmatrix}
			& \partial_j\Phi_l \\
			\partial_j\Phi_l^\dagger
		\end{pmatrix}\otimes\1_2 +\begin{pmatrix}
			& -\partial_0\Phi_l\\
			\partial_0\Phi_l^\dagger
		\end{pmatrix}\otimes\sigma^j,
	\end{split}
\end{equation}
hence
\begin{equation}
	\begin{split}
		\kappa^{-1}\hbox{Tr}&(\Omega_{0j}\Omega_{0j})=-12F_{0j}F_{0j}-2\mathrm{Tr}(W_{0j}W_{0j})-12A_0^2\mathrm{Tr}(\Phi_l^\dagger \Phi_l)-12\mathrm{Tr}(W_0^2\Phi_l^\dagger\Phi_l)\\
		&-4A_j^2\mathrm{Tr}(\Phi_l^\dagger\Phi_l)-4\mathrm{Tr}\left(W_j^2\Phi_l^\dagger \Phi_l\right)-48\mathrm{Tr}(\Phi_l^\dagger\Phi_l)^2 +24A_0\mathrm{Tr}(\Phi_l^\dagger \sigma^3\Phi_l W_0)\\
		&+8A_j\mathrm{Tr}(\Phi_l^\dagger \sigma^3\Phi_lW_j)-4\mathrm{Tr}\left[(\partial_j\Phi_l)^\dagger(\partial_j \Phi_l)\right]-12\mathrm{Tr}\left[(\partial_0\Phi_l)^\dagger(\partial_0\Phi_l)\right]\\
		&-12iA_0\mathrm{Tr}\left[\left((\partial_0\Phi_l)\Phi_l^\dagger-\Phi_l(\partial_0\Phi_l^\dagger)\right)\sigma^3\right]-4iA_j\mathrm{Tr}\left[\left((\partial_j\Phi_l)\Phi_l^\dagger-\Phi_l(\partial_j\Phi_l^\dagger)\right)\sigma^3\right]\\
		&+12i\mathrm{Tr}\left[\left(\Phi_l^\dagger (\partial_0\Phi_l)-(\partial_0\Phi_l^\dagger)\Phi_l\right)W_0\right]+4i\mathrm{Tr}\left[\left(\Phi_l^\dagger (\partial_j\Phi_l)-(\partial_j\Phi_l^\dagger)\Phi_l\right)W_j\right].
	\end{split}
\end{equation}
Moreover,
\begin{equation}
	\begin{split}
		\Omega_{jk}&=-iF_{jk}\begin{pmatrix}
			\sigma^3-\1_2 & \\
			& -\1_2
		\end{pmatrix}\otimes\1_2 - i\begin{pmatrix}
			\00_2 & \\
			& W_{jk}
		\end{pmatrix}\otimes\1_2 +\begin{pmatrix}
			& -\partial_j\Phi_l \\
			\partial_j\Phi_l^\dagger &
		\end{pmatrix}\otimes\sigma^k\\
		&-\begin{pmatrix}
			& -\partial_k\Phi_l \\
			\partial_k\Phi_l^\dagger
		\end{pmatrix}\otimes \sigma^j -2i\varepsilon^{jkl}\begin{pmatrix}
			\Phi_l\Phi_l^\dagger & \\
			& \Phi_l^\dagger \Phi_l 
		\end{pmatrix}\otimes\sigma^l-i\begin{pmatrix}
			& \Phi_lW_j\\
			W_j\Phi_l^\dagger &
		\end{pmatrix}\otimes\sigma^k\\
		&+i\begin{pmatrix}
			& \Phi_lW_k \\
			W_k\Phi_l^\dagger &
		\end{pmatrix}\otimes\sigma^j +i\begin{pmatrix}
			& \sigma^3\Phi_l\\
			\Phi_l^\dagger \sigma^3
		\end{pmatrix}\otimes (A_j\sigma^k-A_k\sigma^j).
	\end{split}
\end{equation}
so that

\begin{equation}
	\begin{split}
		\kappa^{-1}\hbox{Tr}(\Omega_{jk}\Omega_{jk})=&
		-12F_{jk}F_{jk}-2\mathrm{Tr}(W_{jk}W_{jk})
		-96\mathrm{Tr}(\Phi_l^\dagger \Phi_l)^2 \\ &-16\mathrm{Tr}\left[(\partial_j\Phi_l^\dagger)(\partial_j\Phi_l)\right]
		-16\mathrm{Tr}(W_j^2 \Phi_l^\dagger \Phi_l)-16 A_j^2 \mathrm{Tr}(\Phi_l^\dagger \Phi_l) \\
		&+32 A_j\mathrm{Tr}(\Phi_lW_j \Phi_l^\dagger \sigma^3)
		-16i\mathrm{Tr}\left[\left((\partial_j\Phi_l^\dagger)\Phi_l-\Phi_l^\dagger (\partial_j\Phi_l)\right)W_j\right]\\
		&+16iA_j\mathrm{Tr}\left[\left(\Phi_l(\partial_j\Phi_l^\dagger)-(\partial_j\Phi_l)\Phi_l^\dagger\right)\sigma^3\right].
	\end{split}
\end{equation}
Therefore,
\begin{equation}
	\begin{split}
		\kappa^{-1}\hbox{Tr}(\Omega^2)=&
		-12F^2-2\mathrm{Tr}(W^2)
		-24A_\mu A_\mu \mathrm{Tr}(\Phi_l^\dagger \Phi_l) -24\mathrm{Tr}(W_\mu W_\mu \Phi_l^\dagger \Phi_l) \\
		&-192\mathrm{Tr}(\Phi_l^\dagger \Phi_l)^2
		+48A_\mu \mathrm{Tr}\left(\Phi_l^\dagger\sigma^3 \Phi_l W_\mu\right)
		-24\mathrm{Tr}\left[(\partial_\mu \Phi_l)^\dagger(\partial_\mu \Phi_l)\right]
		\\
		&
		-24iA_\mu \mathrm{Tr}\left[\left((\partial_\mu \Phi_l)\Phi_l^\dagger - \Phi_l (\partial_\mu \Phi_l^\dagger)\right) \sigma^3\right] \\ & +24i\mathrm{Tr}\left[\left(\Phi_l^\dagger (\partial_\mu \Phi_l)-(\partial_\mu\Phi_l^\dagger)\Phi_l\right) W_\mu\right],
	\end{split}
\end{equation}
where the summation is performed over Euclidean indices. 

As a result, in the leptonic sector we have
\begin{equation}
	\begin{split}
		a_2 &=\frac{3\kappa}{4\pi^2}\int d^4x \, \mathrm{Tr}(\Phi_l^\dagger \Phi_l), \\
		a_4 & = \frac{\kappa}{48\pi^2}\int d^4x \left[ 6 \left( \mathrm{Tr}(\Phi_l^\dagger\Phi_l)^2  
		-\mathrm{Tr}[(\partial_\mu\Phi_l^\dagger)(\partial_\mu\Phi_l)-A^2\mathrm{Tr}(\Phi_l^\dagger\Phi_l)\right.\right.\\
		&\left.\left. \qquad  -\mathrm{Tr}(W_\mu W_\mu\Phi_l^\dagger \Phi_l)
		-iA_\mu\mathrm{Tr}\left[((\partial_\mu\Phi_l)\Phi_l^\dagger -\Phi_l(\partial_\mu\Phi_l^\dagger) )\sigma^3\right]
		\right.\right.\\ & \left.\left. 
		\qquad +i\mathrm{Tr}\left[(\Phi_l^\dagger (\partial_\mu\Phi_l)-(\partial_\mu\Phi_l^\dagger)\Phi_l )W_\mu\right] 
		+2A_\mu\mathrm{Tr}\left(\Phi_l^\dagger \sigma^3\Phi_lW_\mu\right) \right) \right. \\
		& \left. \qquad +6F^2+\mathrm{Tr}(W^2)+6\varepsilon^{jkl}F_{jk}F_{0l}-3\varepsilon^{jkl}\mathrm{Tr}(W_{jk}W_{0l})\right].
	\end{split}
\end{equation}
Using the parametrization from section \ref{sec:parametrization} we can further write
\begin{equation}
	a_2=\frac{3\kappa}{4\pi^2}(|\Upsilon_e|^2 +|\Upsilon_\nu|^2)\int d^4x |H|^2,
\end{equation}
and
\begin{equation}
	\begin{split}
		a_4=\frac{\kappa}{8\pi^2}\int d^4x & \left[(|\Upsilon_\nu|^4+|\Upsilon_e|^4)|H|^4-(|\Upsilon_\nu|^2+|\Upsilon_e|^2)\mathrm{Tr}|D_\mu H|^2 \right. \\
		&\left.+F^2+\frac{1}{6}\mathrm{Tr}(W^2)+\varepsilon^{jkl}F_{jk}F_{0l}-\frac{1}{2}\varepsilon^{jkl}\mathrm{Tr}(W_{jk}W_{0l})\right].
	\end{split}
\end{equation}
\subsection{Quark sector}
For the quark sector, starting from \eqref{DiracWickq}, we get
\begin{equation}
	\begin{split}
		D_{Q,w}^\dagger D_{Q,w}&=-\Delta_E +2i\left[A_0\begin{pmatrix}
			\sigma^3+\frac{1}{3}\1_2 & \\
			& \frac{1}{3}\1_2
		\end{pmatrix}\otimes \1_2\otimes\1_3 + \begin{pmatrix}
			\00_2 & \\
			& W_0
		\end{pmatrix}\otimes\1_2\otimes\1_3 \right. \\
		&+\left. \1_4\otimes \1_2\otimes G_0 + \begin{pmatrix}
			& \Phi_q\\
			\Phi_q^\dagger &
		\end{pmatrix}\otimes \1_2\otimes\1_3  \right]\partial_0 +2i\left[A_j \begin{pmatrix}
			\sigma^3+\frac{1}{3}\1_2 & \\
			& \frac{1}{3}\1_2
		\end{pmatrix}\otimes\1_2\otimes\1_3 \right. \\
		& \left. +\begin{pmatrix}
			\00_2 & \\
			& W_j
		\end{pmatrix}\otimes\1_2\otimes\1_3+ \1_4\otimes \1_2\otimes G_j + \begin{pmatrix}
			& -i\Phi_q\\
			i\Phi_q^\dagger &
		\end{pmatrix}\otimes \sigma^j\otimes\1_3 \right]\partial_j\\
		&+i(\partial_0A_0)\begin{pmatrix}
			\sigma^3+\frac{1}{3}\1_2 & \\
			&\frac{1}{3}\1_2
		\end{pmatrix}\otimes\1_2\otimes\1_3 +i(\partial_j A_k)\begin{pmatrix}
			\sigma^3+\frac{1}{3}\1_2 & \\
			& \frac{1}{3}\1_2
		\end{pmatrix}\otimes\sigma^j\sigma^k\otimes \1_3\\
		&+A_0^2 \begin{pmatrix}
			\frac{2}{3}\sigma^3 +\frac{10}{9}\1_2 & \\
			& \frac{1}{9}\1_2
		\end{pmatrix}\otimes \1_2\otimes\1_3 +A_jA_k \begin{pmatrix}
			\frac{2}{3}\sigma^3+\frac{10}{9}\1_2 & \\
			& \frac{1}{9}\1_2
		\end{pmatrix}\otimes\sigma^j\sigma^k\otimes\1_3\\
		&+\begin{pmatrix}
			\00_2 &\\
			& W_0^2+i\partial_0W_0
		\end{pmatrix}\otimes\1_2\otimes\1_3 +\begin{pmatrix}
			\00_2 &\\
			& W_jW_k +i\partial_j W_k
		\end{pmatrix}\otimes \sigma^j\sigma^k \otimes \1_3\\
		&+\1_4\otimes\1_2\otimes (G_0^2+i\partial_0G_0)+\1_4\otimes\sigma^j\sigma^k\otimes (G_jG_k+i\partial_jG_k)\\
		&-F_{0j}\begin{pmatrix}
			\sigma^3+\frac{1}{3}\1_2 &\\
			&-\frac{1}{3}\1_2
		\end{pmatrix}\otimes\sigma^j\otimes \1_3+\begin{pmatrix}
			\00_2&\\
			& W_{0j}
		\end{pmatrix}\otimes \sigma^j\otimes \1_3-\begin{pmatrix}
			\1_2 &\\
			&-\1_2
		\end{pmatrix}\otimes \sigma^j \otimes G_{0j}\\
		&+\frac{2}{3}\begin{pmatrix}
			\00_2&\\
			&A_0W_0+A_jW_j
		\end{pmatrix}\otimes\1_2\otimes\1_3+2\begin{pmatrix}
			\sigma^3+\frac{1}{3}\1_2 &\\
			&\frac{1}{3}\1_2
		\end{pmatrix}\otimes\1_2\otimes (A_0G_0+A_jG_j)\\
		&+2\begin{pmatrix}
			\00_2& \\
			&W_0
		\end{pmatrix}\otimes\1_2\otimes G_0 + 2\begin{pmatrix}
			\00_2 &\\
			& W_j
		\end{pmatrix}\otimes \1_2\otimes G_j+i\begin{pmatrix}
			&\partial_0\Phi_q\\
			\partial_0\Phi_q^\dagger
		\end{pmatrix}\otimes\1_2\otimes\1_3 \\
		&+\begin{pmatrix}
			& \partial_j\Phi_q\\
			-\partial_j\Phi_q^\dagger
		\end{pmatrix}\otimes\sigma^j\otimes\1_3 +A_0\begin{pmatrix}
			&\sigma^3\Phi_q+\frac{2}{3}\Phi_q\\
			\Phi_q^\dagger \sigma^3 +\frac{2}{3}\Phi_q^\dagger
		\end{pmatrix}\otimes \1_2\otimes\1_3\\
		&-iA_j\begin{pmatrix}
			& \sigma^3\Phi_q+\frac{2}{3}\Phi_q\\
			-\Phi_q^\dagger \sigma^3-\frac{2}{3}\Phi_q^\dagger&
		\end{pmatrix}\otimes\sigma^j\otimes \1_3+\begin{pmatrix}
			& \Phi_q W_0\\
			W_0\Phi_q^\dagger &
		\end{pmatrix}\otimes\1_2\otimes\1_3\\
		&+i\begin{pmatrix}
			& -\Phi_qW_j\\
			W_j\Phi_q^\dagger &
		\end{pmatrix}\otimes\sigma^j \otimes\1_3 +2\begin{pmatrix}
			& \Phi_q\\
			\Phi_q^\dagger&
		\end{pmatrix}\otimes\1_2\otimes G_0 \\
		&+2i\begin{pmatrix}
			& -\Phi_q\\
			\Phi_q^\dagger
		\end{pmatrix}\otimes \sigma^j\otimes G_j +\begin{pmatrix}
			\Phi_q\Phi_q^\dagger &\\
			&\Phi_q^\dagger \Phi_q
		\end{pmatrix}\otimes \1_2\otimes\1_3.
	\end{split}
\end{equation}
In this case we therefore have
\begin{equation}
	\begin{split}
		E&=\frac{1}{2}F_{jk}\varepsilon^{jkl}\begin{pmatrix}
			\sigma^3+\frac{1}{3}\1_2 &\\
			&\frac{1}{3}\1_2
		\end{pmatrix}\otimes\sigma^l\otimes\1_3 +F_{0j}\begin{pmatrix}
			\sigma^3+\frac{1}{3}\1_2 &\\
			& -\frac{1}{3}\1_2
		\end{pmatrix}\otimes\sigma^j \otimes\1_3 \\
		&+\frac{1}{2}\varepsilon^{jkl}\begin{pmatrix}
			\00_2 &\\
			& W_{jk}
		\end{pmatrix}\otimes\sigma^l\otimes\1_3 -\begin{pmatrix}
			\00_2&\\
			& W_{0j}
		\end{pmatrix}\otimes \sigma^j \otimes \1_3\\
		&+\frac{1}{2}\varepsilon^{jkl}\1_4\otimes \sigma^l \otimes G_{jk} +\begin{pmatrix}
			\1_2 &\\
			&-\1_2
		\end{pmatrix}\otimes\sigma^j\otimes G_{0j}\\
		&+3\begin{pmatrix}
			\Phi_q\Phi_q^\dagger &\\
			& \Phi_q^\dagger \Phi_q
		\end{pmatrix}\otimes\1_2\otimes\1_3.
	\end{split}
\end{equation}
Hence
\begin{equation}
	\hbox{Tr}(E)=36\kappa \mathrm{Tr}(\Phi_q^\dagger \Phi_q).
\end{equation}
Furthermore, we have
\begin{equation}
	\hbox{Tr}(E^2)=\frac{22}{3}F^2+3\mathrm{Tr}(W^2)+4\mathrm{Tr}(G^2)+12\varepsilon^{jkl}F_{jk}F_{0l}-6\varepsilon^{jkl}\mathrm{Tr}(W_{jk}W_{0l})+108\mathrm{Tr}(\Phi_q^\dagger\Phi_q)^2.
\end{equation}
Moreover,
\begin{equation}
	\begin{split}
		\Omega_{0j}&=-F_{0j}\begin{pmatrix}
			\sigma^3 +\frac{1}{3}\1_2 &\\
			&\frac{1}{3}\1_2
		\end{pmatrix}\otimes\1_2\otimes\1_3-i\begin{pmatrix}
			\00_2&\\
			&W_{0j}
		\end{pmatrix}\otimes\1_2\otimes\1_3 -i\1_4\otimes\1_2\otimes G_{0j}\\
		&+iA_0\begin{pmatrix}
			&\sigma^3\Phi_q\\
			\Phi_q^\dagger\sigma^3
		\end{pmatrix}\otimes\sigma^j\otimes\1_3-i\begin{pmatrix}
			&\Phi_qW_0\\
			W_0\Phi_q^\dagger&
		\end{pmatrix}\otimes\sigma^j\otimes\1_3\\
		&-A_j\begin{pmatrix}
			&-\sigma^3\Phi_q\\
			\Phi_q^\dagger \sigma^3
		\end{pmatrix}\otimes\1_2\otimes\1_3 +\begin{pmatrix}
			& -\Phi_q W_j\\
			W_j\Phi_q^\dagger &
		\end{pmatrix}\otimes\1_2\otimes\1_3\\
		&-2i\begin{pmatrix}
			\Phi_q\Phi_q^\dagger & \\
			& -\Phi_q^\dagger \Phi_q
		\end{pmatrix}\otimes\sigma^j\otimes \1_3 +i\begin{pmatrix}
			& \partial_j\Phi_q\\
			\partial_j\Phi_q^\dagger&
		\end{pmatrix}\otimes\1_2\otimes\1_3\\
		& +\begin{pmatrix}
			& -\partial_0\Phi_q\\
			\partial_0\Phi_q^\dagger &
		\end{pmatrix}\otimes\sigma^j\otimes\1_3,
	\end{split}
\end{equation}
hence
\begin{equation}
	\begin{split}
		\kappa^{-1}\hbox{Tr}(\Omega_{0j}\Omega_{0j})&=-\frac{44}{3}F_{0j}F_{0j}-6\mathrm{Tr}(W_{0j}W_{0j})-8\mathrm{Tr}(G_{0j}G_{0j})-36A_0^2\mathrm{Tr}(\Phi_q^\dagger \Phi_q)\\
		&-36\mathrm{Tr}(W_0^2\Phi_q^\dagger \Phi_q)-12A_j^2\mathrm{Tr}(\Phi_q^\dagger \Phi_q)-12\mathrm{Tr}(W_j^2\Phi_q^\dagger \Phi_q)-144\mathrm{Tr}(\Phi_q^\dagger \Phi_q)^2\\
		&+72A_0\mathrm{Tr}(\Phi_q^\dagger \sigma^3\Phi_qW_0)+24A_j\mathrm{Tr}(\Phi_q^\dagger \sigma^3\Phi_qW_j)-12\mathrm{Tr}\left[(\partial_j\Phi_q)^\dagger (\partial_j\Phi_q)\right]\\
		&-36\mathrm{Tr}\left[(\partial_0\Phi_q)^\dagger (\partial_0\Phi_q)\right]-36iA_0\mathrm{Tr}\left[\left((\partial_0\Phi_q)\Phi_q^\dagger - \Phi_q(\partial_0\Phi_q)^\dagger\right)\sigma^3\right]\\
		&-12iA_j\mathrm{Tr}\left[\left((\partial_j\Phi_q)\Phi_q^\dagger -\Phi_q(\partial_j \Phi_q)^\dagger\right)\sigma^3\right]+36i\mathrm{Tr}\left[\left(\Phi_q^\dagger (\partial_0\Phi_q)-(\partial_0\Phi_q^\dagger)\Phi_q\right) W_0\right]\\
		&+12i\mathrm{Tr}\left[\left(\Phi_q^\dagger(\partial_j\Phi_q)-(\partial_j\Phi_q)^\dagger\Phi_q\right)W_j\right].
	\end{split}
\end{equation}
Moreover,
\begin{equation}
	\begin{split}
		\Omega_{jk}&=-iF_{jk}\begin{pmatrix}
			\sigma^3+\frac{1}{3}\1_2 &\\
			& \frac{1}{3}\1_2
		\end{pmatrix}\otimes\1_2\otimes\1_3 -i\begin{pmatrix}
			\00_2&\\
			& W_{jk}
		\end{pmatrix}\otimes\1_2\otimes\1_3 -i\1_4\otimes \1_2\otimes G_{jk}+\\
		&+\begin{pmatrix}
			& -\partial_j \Phi_q\\
			\partial_j \Phi_q^\dagger
		\end{pmatrix}\otimes \sigma^k\otimes\1_3-\begin{pmatrix}
			&-\partial_k\Phi_q\\
			\partial_k\Phi_q^\dagger&
		\end{pmatrix}\otimes\sigma^j\otimes\1_3\\
		&-2i\varepsilon^{jkl}\begin{pmatrix}
			\Phi_q\Phi_q^\dagger & \\
			& \Phi_q^\dagger \Phi_q
		\end{pmatrix}\otimes\sigma^l\otimes\1_3-i\begin{pmatrix}
			& \Phi_q W_j\\
			W_j\Phi_q^\dagger&
		\end{pmatrix}\otimes \sigma^k\otimes\1_3\\
		& +i\begin{pmatrix}
			& \Phi_q W_k\\
			W_k\Phi_q^\dagger &
		\end{pmatrix}\otimes\sigma^j\otimes\1_3 +i \begin{pmatrix}
			& \sigma^3\Phi_q\\
			\Phi_q^\dagger \sigma^3 &
		\end{pmatrix}\otimes (A_j\sigma^k-A_k\sigma^j)\otimes\1_3,
	\end{split}
\end{equation}
and
\begin{equation}
	\begin{split}
		\kappa^{-1}\hbox{Tr}(\Omega_{jk}\Omega_{jk})&=-\frac{44}{3}F_{jk}F_{jk}-6\mathrm{Tr}(W_{jk}W_{jk})-8\mathrm{Tr}(G_{jk}G_{jk})-288\mathrm{Tr}(\Phi_q^\dagger \Phi_q)^2\\
		&-48\mathrm{Tr}\left[(\partial_j\Phi_q)^\dagger (\partial_j \Phi_q)\right]-48\mathrm{Tr}(W_j^2\Phi_q^\dagger \Phi_q)-48A_j^2 \mathrm{Tr}(\Phi_q^\dagger \Phi_q)\\
		&+96A_j \mathrm{Tr}(\Phi_q W_j \Phi_q^\dagger \sigma^3)-48i\mathrm{Tr}\left[\left((\partial_j\Phi_q)^\dagger \Phi_q-\Phi_q^\dagger (\partial_j \Phi_q)\right)W_j\right]\\
		&+48iA_j \mathrm{Tr}\left[\left(\Phi_q(\partial_j \Phi_q)^\dagger-(\partial_j\Phi_q)\Phi_q^\dagger\right)\sigma^3\right].
	\end{split}
\end{equation}
Therefore
\begin{equation}
	\begin{split}
		\kappa^{-1}\hbox{Tr}(\Omega^2)&= - 2\left(\frac{22}{3}F^2 + 3\mathrm{Tr}(W^2)+4\mathrm{Tr}(G^2)\right)-576\mathrm{Tr}(\Phi_q^\dagger \Phi_q)^2\\
		&-72\mathrm{Tr}\left[(\partial_\mu \Phi_q)^\dagger (\partial_\mu \Phi_q)\right]-72\mathrm{Tr}\left(W_\mu W_\mu \Phi_q^\dagger \Phi_q\right)-72A_\mu A_\mu \mathrm{Tr}\left(\Phi_q^\dagger \Phi_q\right)\\
		&+144 A_\mu \mathrm{Tr}\left(\Phi_q^\dagger\sigma^3\Phi_q W_\mu\right) -72iA_\mu \mathrm{Tr}\left[\left((\partial_\mu \Phi_q)\Phi_q^\dagger-\Phi_q(\partial_\mu \Phi_q)^\dagger\right)\sigma^3\right]\\
		&+72i\mathrm{Tr}\left[\left(\Phi_q^\dagger (\partial_\mu \Phi_q)-(\partial_\mu \Phi_q)^\dagger \Phi_q\right)W_\mu\right].
	\end{split}
\end{equation}
Expressing the coefficients $a_2$ and $a_4$ as in the section \ref{sec:parametrization} we can further write
\begin{equation}
	a_2=\frac{3\kappa}{4\pi^2}(|\Upsilon_e|^2 +|\Upsilon_\nu|^2)\int d^4x 3|H|^2,
\end{equation}
and
\begin{equation}
	\begin{split}
		a_4=& \frac{\kappa}{8\pi^2}\int d^4x  \biggl[(3|\Upsilon_\nu|^4+3|\Upsilon_e|^4)|H|^4-(3|\Upsilon_\nu|^2+3|\Upsilon_e|^2)\mathrm{Tr}|D_\mu H|^2  \\
		&+\frac{11}{9}F^2+\frac{1}{2}\mathrm{Tr}(W^2)+\frac{2}{3}\mathrm{Tr}(G^2)+3\varepsilon^{jkl}F_{jk}F_{0l}-\frac{3}{2}\varepsilon^{jkl}\mathrm{Tr}(W_{jk}W_{0l})\biggr].
	\end{split}
\end{equation}

\acknowledgments
AB acknowledges the hospitality of the Department of Mathematics of the Indiana University Bloomington during the Fulbright Junior Research Award scholarship funded by the Polish-US Fulbright Commission. 

\noindent AB was partially supported by Faculty of Physics, Astronomy and Applied Computer Science of the Jagiellonian University under the MNS scheme: N17/MNS/000010.

\noindent AS and PZ acknowledge the support by NCN Grant 2020/37/B/ST1/01540.

\end{document}